\documentclass[fleqn]{article}
\usepackage{mcite}
\usepackage{graphicx,rotating}
\graphicspath{{/home/critten/photon97/}}
\DeclareGraphicsExtensions{.eps,.ps,.eps.gz,.ps.gz}
\newcommand{\gev}{{\rm Ge}\kern-1.pt{\rm V}}
\newcommand{\gevsq}{\mbox{$\mathrm{{\rm Ge}\kern-1.pt{\rm V}}^2$}}

\newcommand{\rhoz}{\mbox{$\rho^0$}}
\newcommand{\gptovp}{\mbox{$\scriptstyle{{\gamma^*{\rm p}}} \rightarrow {\rm {\rho}p}$}}\newcommand{\jpsi}{\mbox{$J/\psi$}}
\newcommand{\xbj}{\mbox{$x$}}
\newcommand{\qsq}{\mbox{$Q^2$}}
\newcommand{\kk}{\mbox{\bf k}}
\newcommand{\kkp}{\mbox{{\bf k}$'$}}
\newcommand{\pp}{\mbox{\bf P}}
\newcommand{\ppp}{\mbox{{\bf P}$'$}}
\newcommand{\qq}{\mbox{\bf q}}
\newcommand{\vv}{\mbox{\bf V}}

\newcommand{\phih}{\mbox{${\phi}_h$}}
\newcommand{\Phih}{\mbox{${\Phi}_h$}}
\newcommand{\D}{\mbox{\rm{d}}}

\newcommand{\thetah}{\mbox{${\theta}_h$}}

\newcommand{\fsq}{\mbox{$F_2$}}

\newcommand{\qqbar}{\mbox{$q\bar{q}$}}

\newcommand{\Pma}{I\!\!P}

\def\lsim{\mathrel{\rlap{\lower4pt\hbox{\hskip1pt$\sim$}}
    \raise2pt\hbox{$<$}}} %less than or approx. symbol
\def\gsim{\mathrel{\rlap{\lower4pt\hbox{\hskip1pt$\sim$}}
    \raise2pt\hbox{$>$}}} %greater than or approx. symbol
\begin{document}
%\draft
%\preprint{HEP/123-qed}
\thispagestyle{empty}
%\hfill BONN-HE-xxx\\*[1cm]
\begin{center}
\begin{LARGE}
{\bf Scale Issues in High-Energy\\ Diffractive Vector-Meson Production}\\*[1cm]
\end{LARGE}
\begin{large}
{\sc J.A.~Crittenden}\\*[3mm]
{\it Physikalisches 
Institut, Universit\"at Bonn\\ Nu{\ss}allee 12, 53115 Bonn, Germany\\
E-mail : crittenden@desy.de}\\*[1cm]
\end{large}
\end{center}
%\date{\today}
%\maketitle
%\begin{abstract}
\noindent
Recent measurements of diffractive vector-meson production with the general-purpose  detectors H1 and ZEUS in electron-proton interactions with 300 GeV center-of-mass energy at the HERA accelerator complex have stimulated great interest in the question of whether perturbative QCD (pQCD) can provide an accurate description of such diffractive processes. The high flux of quasi-real photons from the 27.5 GeV electron beam has allowed high-statistics studies of vector-meson photoproduction to be compared to the deeply inelastic processes at high photon virtuality $Q^2 \gg \mathrm{\Lambda}_{\mathrm{QCD}}^2$. Special-purpose electron detectors at small scattering angle have been used to identify a transition region, $0.2 < \qsq < 2\;{\mathrm {GeV}}^2$, where the pQCD prediction for the dominance of point-like configurations of q$\overline{\mathrm{q}}$-pairs is verified. 
Data samples characterized by high momentum transfer to the 820 GeV initial-state proton, $1 < |t| < 13\;{\mathrm{GeV}}^2$, have now become available, and the results lend support to the proposition that pQCD calculations should accurately describe the inelastic vector-meson production processes in high-$|t|$ photoproduction. Measurements of the energy dependence and the $|t|$-dependence of the elastic cross sections for $\rho, \omega, \phi$, and $\jpsi$ production have provided tests of the proposal that the mass (or size) of the vector meson provides a third scaling variable for studying the transition to the domain of pQCD applicability. A preliminary result on elastic $\Upsilon$ production has also been announced by the ZEUS collaboration. These issues of scale will be discussed in the context of topical phenomenological models, comparing the HERA results and those from fixed-target experiments at lower energy to the wealth of calculations available in the contemporary literature.\\*[3cm]
Contribution to the proceedings of the LISHEP International School on High-Energy Physics, Rio de Janeiro, Brazil, 16-20 February 1998
\clearpage
\pagenumbering{roman}
\setcounter{page}{2}
\tableofcontents
\clearpage
\pagenumbering{arabic}
\setcounter{page}{1}
\section{Introduction}
\label{sec:intro}
The successful operation of the HERA electron-proton collider\footnote{The results discussed in this talk were obtained from data recorded since 1994, a period during which HERA operated a positron beam in collision with the proton beam. In our treatment of high-energy diffraction we will ignore effects associated with the sign of the initial-state lepton, and refer to it as ``electron'' in generic fashion. HERA is scheduled to run with an $e^-$ beam in 1998.} at the German national laboratory DESY (Deutsches Elektronen-Synchrotron) has opened a new era 
of experimental investigation into diffractive vector-meson photo- and leptoproduction. The ability of the HERA accelerator physicists to bring 27.5 GeV electrons into collision with 820 GeV protons results in a kinematic limit of 300 GeV for the center-of-mass energy in the resulting photon-proton interactions, equivalent to that of a 50 TeV photon beam incident on a stationary proton target. This extension of the accessible energy range by more than an order of magnitude over previous experiments is of particular importance for studies of diffractive processes, since sensitivity to their characteristically weak energy dependence is thus enhanced, and since energy-dependent background processes are suppressed. The high flux of quasi-real photons from the electron beam further provides a copious source of photoproduced reactions which can be compared to deeply inelastic processes, a comparison spanning more than eleven orders of magnitude in photon virtuality.{\kern-4.pt}~\footnote{Although the initial state for all interactions studied at HERA consists of an electron and a proton, the HERA collaborations ZEUS and H1 have adopted a convention of referring as ``photoproduction'' to neutral-current reactions in which the final-state electron escapes detection in the central detector by exiting via the beam pipe at small scattering angle, since in this case the typical values for the photon virtuality lie several orders of magnitude below the squared hadronic mass scale.}

\begin{figure}[htbp]
\vspace*{-5mm}
\begin{minipage}[t]{0.48\textwidth}
\raisebox{5mm}{\includegraphics[width=0.95\textwidth]{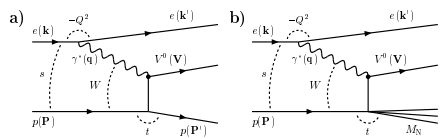}}
\end{minipage}
\hfill
\begin{minipage}[b]{0.48\textwidth}
\caption
{
\label{fig:newvector3}
Schematic diagrams for diffractive vector meson production in electron-proton interactions, distinguishing a)~elastic and b)~proton dissociative processes. See text for definitions of the variables and four-vectors
\hspace*{\fill}
}
\end{minipage}
\vspace*{-5mm}
\end{figure}

%This article concerns the diffractive production of vector mesons, where the momentum transfer to the proton is exponentially suppressed. 
This article concerns the diffractive production of vector mesons, a subject with a rich history, having been studied since the 1960s. Bauer et al.~\cite{rmp_50_261} published a comprehensive review of the subject in 1978. In 1984, Holmes et al.~\cite{arnps_35_397} reviewed the status of investigations of {\jpsi} photoproduction. A summary of the investigations at HERA prior to 1997 and of the recent theoretical work stimulated by these measurements was published last year~\cite{stmp_140}. These processes are characterized by forward scattering of the proton or of the associated hadronic system in case of proton dissociation. Figure~\ref{fig:newvector3} 
distinguishes the processes we refer to as elastic and proton dissociative.
It also illustrates our definitions of the kinematic variables defined in terms of the four-vectors, for which we specify here the ranges relevant to the results covered in this article:\\[2mm]
$s$, the center-of-mass energy of the electron-proton system,
\begin{eqnarray*}
s&=&(\kk + \pp)^2,\hspace{5mm}\sqrt{s} = 300 \; \gev ,
\end{eqnarray*}
$W$, the center-of-mass energy of the photon-proton system,
\begin{eqnarray*}
W^2&=&(\qq + \pp)^2,\hspace{5mm}20<W<240\;\gev ,
\end{eqnarray*}
$t$, the squared momentum transfer to the proton,
\begin{eqnarray*}
t & \equiv & (\pp - \ppp)^2 \, = \, (\vv - \qq)^2,\hspace{5mm}-13<t<-|t|_{\rm min},
\end{eqnarray*}
where
\begin{eqnarray*}
|t|_{\rm min} &=& m_{\rm p}^2 \, \frac{(M_{\rm V}^2 + \qsq)^2}{W^4},
\end{eqnarray*}
with $M_{\rm V}$ indicating the vector-meson mass, $m_{\rm p}$ the proton mass, and the photon virtuality designated by \mbox{${\qsq}\,{\equiv}\,-q^2\,=\,-(\kk - \kkp)^2$}.

The characterization of the process in Fig.~\ref{fig:newvector3}~a) as ``elastic'' is a misnomer, since rest mass is created in the reaction.  The creation of rest mass results in a kinematically required minimum momentum transfer to the proton, $|t|_{\rm min}$. At HERA energies this minimum is negligible ($|t|_{\rm min} \approx 10^{-4}\; \gevsq$) with respect to the average value of $|t|$, motivating the use of the term elastic to distinguish this process from those in which the proton dissociates. For investigations at HERA where the final-state proton is not detected, the dissociative process typically represents a 10-20\% background correction for the extraction of the elastic cross section for $|t|<1\;\gevsq$, but dominates at high $|t|$.

Phenomenological interest in recent years has been spurred by indications that these diffractive processes are calculable in the context of perturbative quantum chromodynamics. The purpose of this article is to discuss the proposition that the hard scale necessary to the perturbative expansion may be given by either {\qsq}, $M^2_{\rm V}$, or $|t|$.

\section{Signatures for Perturbative QCD}
\label{sec:signatures}
Several types of experimental evidence for the applicability of perturbative QCD (pQCD) calculations serve our discussion of scale issues. Some arguments will be based on the consistency of the measurements with specific predictions of pQCD calculations. Others will refer to the absence of effects known to be of nonperturbative origin. This section serves to introduce these arguments in order to facilitate the subsequent presentation of recent experimental results.
\subsection*{Energy Dependence of Elastic Cross Sections}
\begin{figure}[htbp]
%\vspace*{-5mm}
%\begin{minipage}[t]{0.48\textwidth}
\includegraphics[width=0.8\textwidth]{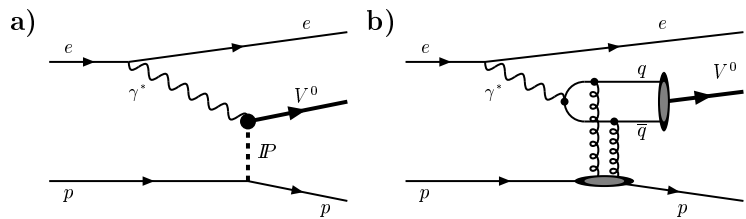}
%\end{minipage}
%\hfill
%\begin{minipage}[b]{0.48\textwidth}
\caption
{
\label{fig:vmdiag2}
Diagrams illustrating two types of elastic diffractive vector-meson production mechanisms: a)~Pomeron exchange, calculated in the context of Regge phenomenology, and b)~exchange of a gluon pair, calculated via perturbative QCD
\hspace*{\fill}
}
%\end{minipage}
%\vspace*{-5mm}
\end{figure}

The initial observation of elastic {\rhoz} photoproduction at HERA with a cross section comparable to that measured at an order-of-magnitude lower energy, and with an exponentially falling $t$-dependence, served as convincing evidence for the diffractive nature of the observed reaction~\cite{zfp_69_39}. An amazingly successful phenomenological description of total hadronic cross sections (and of forward elastic cross sections via the optical theorem) had been obtained by applying Regge theory and postulating a mixture of Pomeron and Reggeon exchange~\cite{pr_54_1,*pl_296_227}. This parametrization of the energy dependence yields agreement with measurements over four orders of magnitude in center-of-mass energy, the Pomeron contribution dominating above 50 \gev. According to this prescription, diagrammed in Fig.~\ref{fig:vmdiag2}~a), 
the differential $\gamma p \rightarrow \rhoz p$ cross section is described~\cite{pl_348_213,*pl_397_317} as
\begin{eqnarray*}
\label{eq:ebt}
\frac{\D\sigma}{\D{|t|}} &\propto& 
e^{-b_0 |t|} \left( \frac{s}{s_0} \right)^{2({\alpha_{\Pma}(t)}-1)},
\hspace*{1cm}{\alpha_{\Pma}(t)=1.08\,+\,0.25\,t}.
\end{eqnarray*}
The attendant logarithmic decrease of the exponential $t$-slope with energy~\cite{jetp_14_478} is discussed below. The energy dependence of the forward cross section is then
\begin{eqnarray*}
\left. \frac{\D\sigma}{\D{|t|}} \right|_{t=0}&\propto& \left( \frac{s}{s_0} \right)^{2({\alpha_{\Pma}(0)}-1)}\;\propto\;W^{4\epsilon}\;=\;W^{0.32}.
\end{eqnarray*}
This weak power dependence on energy proved to give a good description of the HERA {\rhoz} photoproduction cross section, the effective exponent reduced to $\epsilon \approx 0.05$, owing to the integration over the $t$-dependence in the range covered by the data.

\begin{figure}[htbp]
 \begin{center}
  \begin{tabular}{cc}
   \parbox[b]{0.58\textwidth}{\includegraphics[width=0.58\textwidth, bb= 3 18 524 686, clip=]{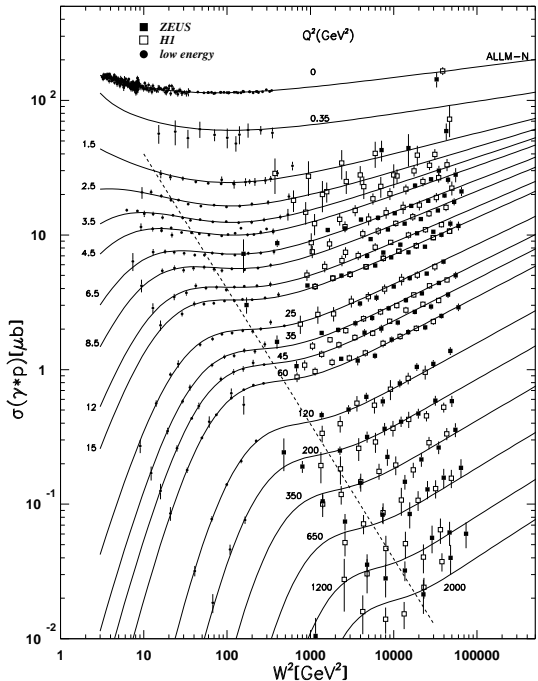}
   \vskip -3mm
   \caption{
   \label{fig:sigtot}
The value of the total
photon--proton cross section $\sigma({{\gamma^*}\rm p})$ as a function of
the squared center-of-mass energy for various values of {\qsq}
as measured by the ZEUS~\protect\cite{zfp_72_399} and H1~\protect\cite{np_470_3} collaborations and by
the NMC collaboration~\protect\cite{pl_364_107} at CERN. The curves represent
calculations using the ALLM proton structure
function par\-a\-me\-tri\-za\-tions~\protect\cite{pl_269_465,*marcusthesis}.
The dashed line connects points where {\xbj}=0.1
}
}
&
\noindent
\parbox[b]{0.38\textwidth}{
The curves result from structure function par\-a\-me\-tri\-za\-tions which cov\-er the trans\-i\-tion through the hadronic mass scale region in photon virtuality~\cite{pl_269_465,*marcusthesis}. Further information on these cross sections and their interpretation can obtained from the talk by A.~Bornheim. Here we point out first the interesting threshold behavior for ${\xbj}>0.06$, where the proton diameter exceeds the virtual photon lifetime in the the proton rest frame. The origin of the steep energy dependence at high {\qsq} and high $W$, the HERA discovery, is clearly in need of an independent explanation. The cross section rises much more steeply than that for quasi-real photons, where the energy dependence is comparable to that of the total had\-ron\-ic cross sections. The behavior of these cross sections at energies higher than those reached at HERA remains an open and intriguing question.{\kern-4.pt}~\footnotemark \hspace*{\fill}
}
   \end{tabular}
  \end{center}
\end{figure}

The calculations of diffractive vector-meson production based on perturbative QCD define the exchanged object as a colorless two-gluon state~\cite{pr_12_163, *prl_34_1286, *pr_14_246, *pr_15_2503}(see Fig.~\ref{fig:vmdiag2}~b)). Ryskin~\cite{zfp_57_89} calculated a cross section for {\jpsi} photoproduction proportional to the square of the gluon density. Brodsky et al.~\cite{pr_50_3134} pointed out that the point-like configurations of a {\qqbar} pair expected at high photon virtuality must be longitudinally polarized, and calculated the diffractive {\rhoz} production cross section for longitudinally polarized photons, also obtaining a dependence on the square of the gluon density. These calculations result in a much steeper energy dependence than described above for the total hadronic cross sections, since the gluon density has been measured by the HERA experiments~\cite{zfp_72_399,np_470_3} to rise steeply with decreasing {\xbj} for ${\xbj}={\qsq}/W^2 \ll 1$. To illustrate this point we show in Fig.~\ref{fig:sigtot} the total $\gamma^* p$ cross section derived from the HERA structure function measurements:
\begin{eqnarray*}
\label{eq:virtot}
\sigma({{\gamma^*}\rm p}) &\equiv& \sigma_{\rm T}({{{\gamma^*}\rm p}}) + \sigma_{\rm L}({{{\gamma^*}\rm p}}) \, \approx \, \frac{4\pi^2\alpha}{\qsq (1-\xbj)} {\fsq}({\xbj},\qsq),\hspace*{5mm}({\xbj}^2 m_p^2 \ll {\qsq}).
\end{eqnarray*}
\footnotetext{The neutrino-nucleon total cross section at low energy is another example of a steeply rising cross section, proportional to $W^2$. This energy dependence was successfully described by the Fermi theory of weak interactions for four decades until the Fermi theory was replaced by the electroweak Standard Model, which is now reaching a similar age.}

\subsection*{Relative Contributions from $\sigma_{\rm L}$ and $\sigma_{\rm T}$}
As mentioned above, QCD considerations lead directly to the expectation that the longitudinal cross section will dominate at high {\qsq}~\cite{pr_50_3134, pr_56_2982}. Measurements of the ratio $R=\sigma_{\rm L}/\sigma_{\rm T}$ provide a direct test.

\subsection*{Dependence on the Momentum Transfer to the Proton}
The exponential slope of elastic cross sections is directly related to the sizes of the scattering objects: $b\;\propto\;R^2_{\mathrm p}\,+\,R^2_{\mathrm {VM}}$. Since QCD predicts the size of the {\qqbar} pair to decrease with increasing photon virtuality, this slope is of particular interest. Brodsky et al.~\cite{pr_50_3134} emphasized their prediction of a flavor-independent slope at high {\qsq}, the slope determined by the two-gluon form factor alone. 

The hadronic cross sections show an energy dependence in this slope,
called 'shrinkage' since the slope steepens with energy:
\begin{eqnarray}
\label{eq:b0}
\frac{\D\sigma}{\D{|t|}} &\propto& e^{-b |t|} \hspace*{4mm} b\;=\;b_0\,+\, 2\,{\alpha_{\Pma}'} \ln{\frac{s}{s_0}}, \hspace{5mm}\alpha_{\Pma}'=0.25\;{\gev}^{-2}.
\end{eqnarray}
Such an effect results from the analyticity and unitarity properties
of the asymptotic elastic scattering amplitude~\cite{jetp_14_478}, and
is unrelated to a QCD description of the cross section. Thus the absence of shrinkage is a positive indicator for the validity of a pQCD approach.

\subsection*{Skewing in the ${\rhoz}\rightarrow \pi^+\pi^-$ Mass Spectrum}
The amplitude for the diffractive production of the {\rhoz} meson and its decay to a pair of pions interferes with that for the continuum production of a pion pair~\cite{prl_5_278}. The resulting skewing of the dipion mass resonance has been extensively studied and discussed~\cite{rmp_50_261,nc_34_1644,*pl_19_702,*pr_149_1172,*pr_9_126}. While the effect is interesting, and may allow the extraction of the total pion-proton cross section at an energy much higher than that at which it has been directly measured, this skewing is irrelevant to QCD considerations. 

\subsection*{Vector Meson Production Ratios}
In 1978 Bauer et al.~\cite{rmp_50_261} referred to a ``serious weakness'' in the understanding of the ratio of $\phi$ and {\rhoz} photoproduction cross sections, citing a factor of two suppression relative to the expectation from Vector Dominance. Assuming a point-like electromagnetic coupling to the valence quark content of the vector mesons, the ratios are given by
\begin{tabular}[t]{c@{\hspace{3mm} : \hspace{3mm}}c@{\hspace{3mm} : \hspace{3mm}}c@{\hspace{3mm} : \hspace{3mm}}c@{\hspace{3mm} : \hspace{3mm}}c}
%\begin{tabular}[t]{ccccc}
\rhoz & $\omega$ & $\phi$ & \jpsi & $\Upsilon$ \\
9 & 1 & 2 & 8 & 2.
\end{tabular}
\linebreak
The rate for {\jpsi} photoproduction was found in 1975 to be three orders of magnitude lower than this value, excluding Vector Dominance as a sufficient description of vector-meson photoproduction.{\kern-4.pt}\footnote{I thank B.~Kopeliovich for pointing out that for models in which a flavor-dependent suppression mechanism applies to the vector-meson-proton cross section, Vector Dominance remains a viable prescription for the photon-vector-meson transition.} Thus we will consider the production ratios to serve as an indicator for point-like, perturbative, couplings.\\

These are the physical phenomena which we will use in the following discussion to address the question of applicability of perturbative QCD calculations to diffractive vector-meson production. At each step we will discuss the issue of hard scale, and whether it can be given by photon virtuality, momentum transfer at the proton vertex, or the mass of the vector meson. The status of the experimental results from the H1 and ZEUS collaborations is that at the time of the 
International Europhysics Conference on High-Energy Physics in Jerusalem last August~\cite{jerusalem}.

\section{Energy Dependence of Elastic Cross Sections}
\subsection{Elastic Vector-Meson Photoproduction}
\label{sec:phpvmxsec}
\begin{figure}[htbp]
\begin{center}
\begin{tabular}{cc}
\parbox[b]{0.58\textwidth}{
\raisebox{-1mm}{\includegraphics[width=0.58\textwidth, bb= 0 0 568 627, clip=]{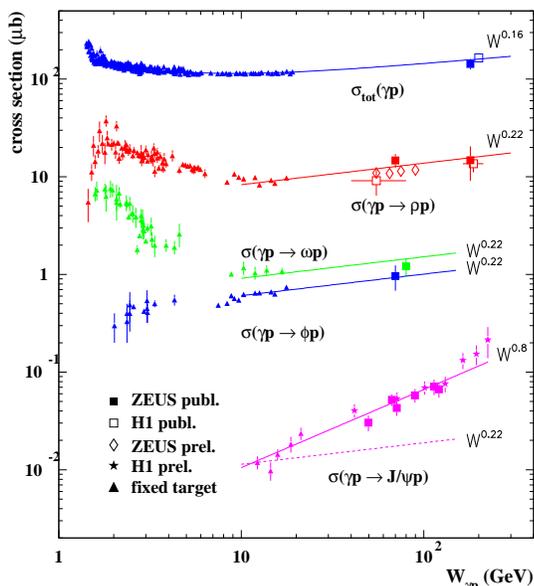}}
\vskip -4mm
\caption
{
\label{fig:phpvm}
Measurements of the elastic photoproduction cross sections for {\rhoz}, $\omega$, $\phi$, and {\jpsi} mesons versus energy, compared to the total $\gamma p$ cross section. The lines indicate the slopes of two types of power-law dependence \hspace*{\fill}
}
}
&
\parbox[b]{0.37\textwidth}{
energy. The resonance contributions to the {\rhoz} and $\omega$ cross sections at low energy are clearly observed. Their absence in $\phi$ production has been used to identify diffractive $\phi$ production as a process particularly suited to the study of Pomeron exchange. The H1 collaboration have used their lead/scintillating-fiber calorimeter to obtain a sample of {\jpsi} mesons boosted in the direction of the initial-state electron momentum, corresponding to values for $W$ greater than 200 {\gev}~\cite{j242, *west}. The steeper energy dependence for {\jpsi} photoproduction far beyond the threshold region is striking. Clearly Pomeron exchange yields an insufficient \hfill description of this process,
}
\end{tabular}
\end{center}
\end{figure}

\begin{figure}[htbp]
\begin{center}
\begin{tabular}{cc}
\parbox[b]{0.58\textwidth}{
\includegraphics[width=0.58\textwidth, bb= 0 34 540 423, clip=]{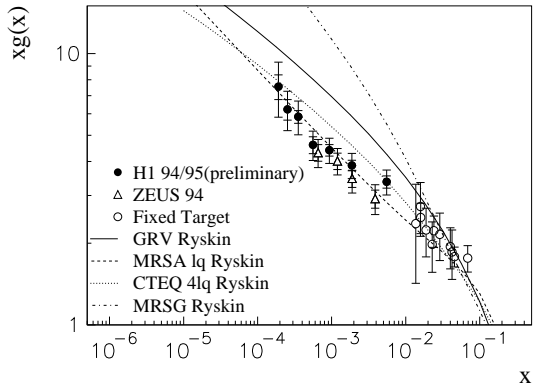}
}&
\parbox[b]{0.37\textwidth}{
\caption
{
\label{fig:h1_xglu}
HERA measurements of the elastic {\jpsi} photopro\-duction cross section
compared to those at lower energy
and to pQCD calculations employing various gluon density
par\-a\-me\-tri\-za\-tions~\protect\cite{j242, *west} \hspace*{\fill}
}
}
\end{tabular}
\end{center}
\end{figure}

Figure~\ref{fig:phpvm} 
shows the measurements presently available for the elastic photoproduction of {\rhoz}, $\omega$, $\phi$, and {\jpsi} mesons, comparing these cross sections to the total $\gamma p$ cross section, of which elastic {\rhoz} production accounts for about 7\% at \hfill high 
\vskip -4mm
\vskip -6mm \noindent
despite the validity of the same argument as was used for the $\phi$ meson. Since the ranges covered in both {\qsq} and $t$ lie far below the hadronic mass scale, this measurement indicates that the {\jpsi} mass alone accounts for the observed consistency with pQCD calculations of the energy dependence. The predictions of such calculations using various gluon density parametrizations are shown in Fig.~\ref{fig:h1_xglu}. The quality of the measurements at high energy clearly allow one to discriminate among the shapes of the parametrizations.
\vskip -5mm

\begin{figure}[htbp]
\vspace*{-3mm}
\begin{center}
\begin{tabular}{cc}
\parbox[t]{0.6\textwidth}{
\raisebox{-2mm}{\includegraphics[width=0.6\textwidth, bb= 0 10 512 544, clip=]{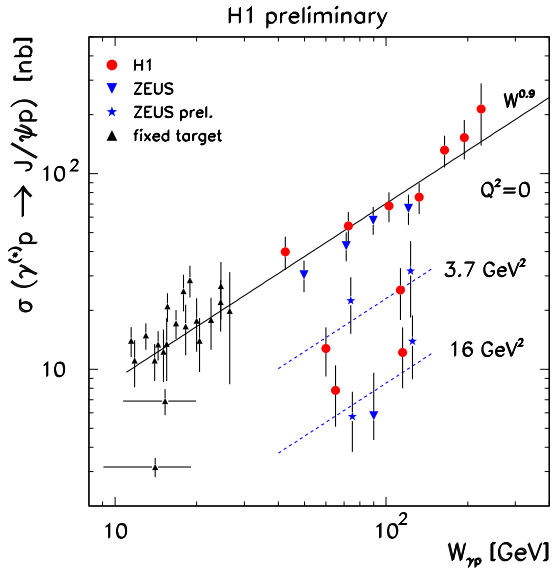}}
}&
\parbox[b]{0.37\textwidth}{
\caption
{
\label{fig:phppsi_h1}
Comparison of the elastic {\jpsi} photoproduction cross section to those measured at
high {\qsq} by the ZEUS and H1 collaborations
}
}
\end{tabular}
\end{center}
\end{figure}

Fig.~\ref{fig:phppsi_h1} compares the measurements of {\jpsi} photoproduction to those at higher values of {\qsq}. The energy dependence at high {\qsq} is consistent with that observed for photoproduction, but statistical limitations prevent conclusive distinctions between the various calculations presently available~\cite{pr_57_512, zfp_76_231, hep96_05_208}, which differ significantly in their predictions for the {\qsq}-dependence.
\vskip -5mm

\begin{figure}[htbp]
%\vspace*{-5mm}
\begin{minipage}[t]{0.62\textwidth}
\raisebox{-1mm}{\includegraphics[width=\textwidth, bb=20 38 513 260, clip=]{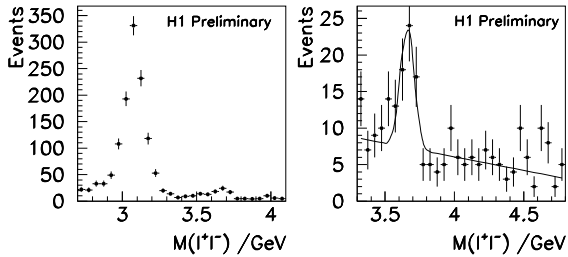}}
\end{minipage}
\hfill
\begin{minipage}[b]{0.35\textwidth}
\caption
{
\label{fig:psi2s_signal}
Signal for the photoproduction of $\psi$(2s) reported by the H1 collaboration~\protect\cite{dr_97_228}
}
\end{minipage}
\vspace*{-2mm}
\end{figure}

The H1 and ZEUS experiments will be presenting further results on the diffractive production of heavy quarkonia in the near future. Figure~\ref{fig:psi2s_signal} shows the $\psi$(2S) signal reported by H1 last year~\cite{dr_97_228}. The elastic cross section derived at an average center-of-mass energy of 80 {\gev}, $17.9 \pm 2.8 \mathrm{(stat)} \pm 2.7 \mathrm{(sys)} \pm 1.4 \mathrm{(BR)}$, is consistent with the prediction of the color dipole model~\cite{hep97_12_469,*zfp_75_71,*pr_44_3466,*pl_309_179}. This model predicts particularly striking {\qsq}-dependent effects in the diffractive production of the radially excited states.

\begin{figure}[htbp]
\begin{center}
\begin{tabular}{lr}
\parbox[b]{0.35\textwidth}{\hspace*{3mm}
The ZEUS collaboration have used a combined data sample from the data-taking periods 1995-97 to extract a dimuon invariant mass spectrum which exhibits clear peaks at the {\jpsi} and {$\psi$(2S)} mass values, as well as about thirty events in the $\Upsilon$ region~\cite{hep97_10_013}, as shown in Fig.~\ref{fig:upsilon}. The exponentially falling background is described by Bethe-Heitler lepton pair production~\cite{np_229_347,*baranov}. The mass resolution is insufficient to resolve the ${\Upsilon}$,${\Upsilon}'$, and ${\Upsilon}''$.
They report a result for the combined elastic production cross section in the energy range $80<W<280\;\gev$ which is about 1\% of the elastic {\jpsi} photoproduction cross section.
}&
\parbox[b]{0.55\textwidth}{
\includegraphics[width=0.55\textwidth, bb= 48 161 530 676, clip=]{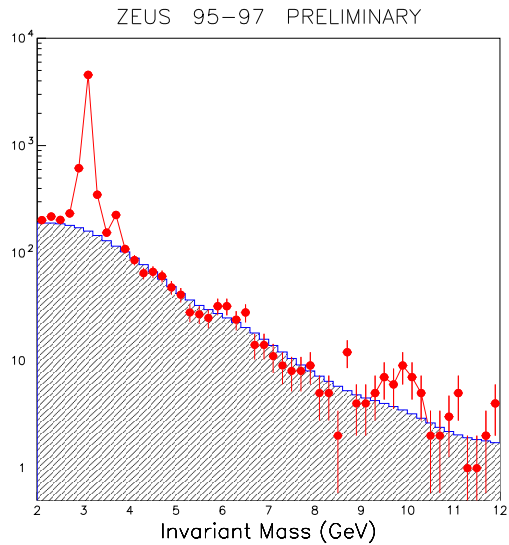}
\caption
{
\label{fig:upsilon}
Dimuon mass spectrum showing {\jpsi}, $\psi$(2S), and $\Upsilon$ peaks presented by the ZEUS collaboration~\protect\cite{hep97_10_013}.
}
}
\end{tabular}
\end{center}
\end{figure}
%\clearpage

\vspace*{-4mm}
\subsection{Production of {\rhoz} Mesons in Deep Inelastic Scattering}
\label{sec:rhoprod}
\begin{figure}[htbp]
\vspace*{-1mm}
\begin{minipage}[t]{0.48\textwidth}
\includegraphics[width=\textwidth, bb=0 22 485 574, clip=]{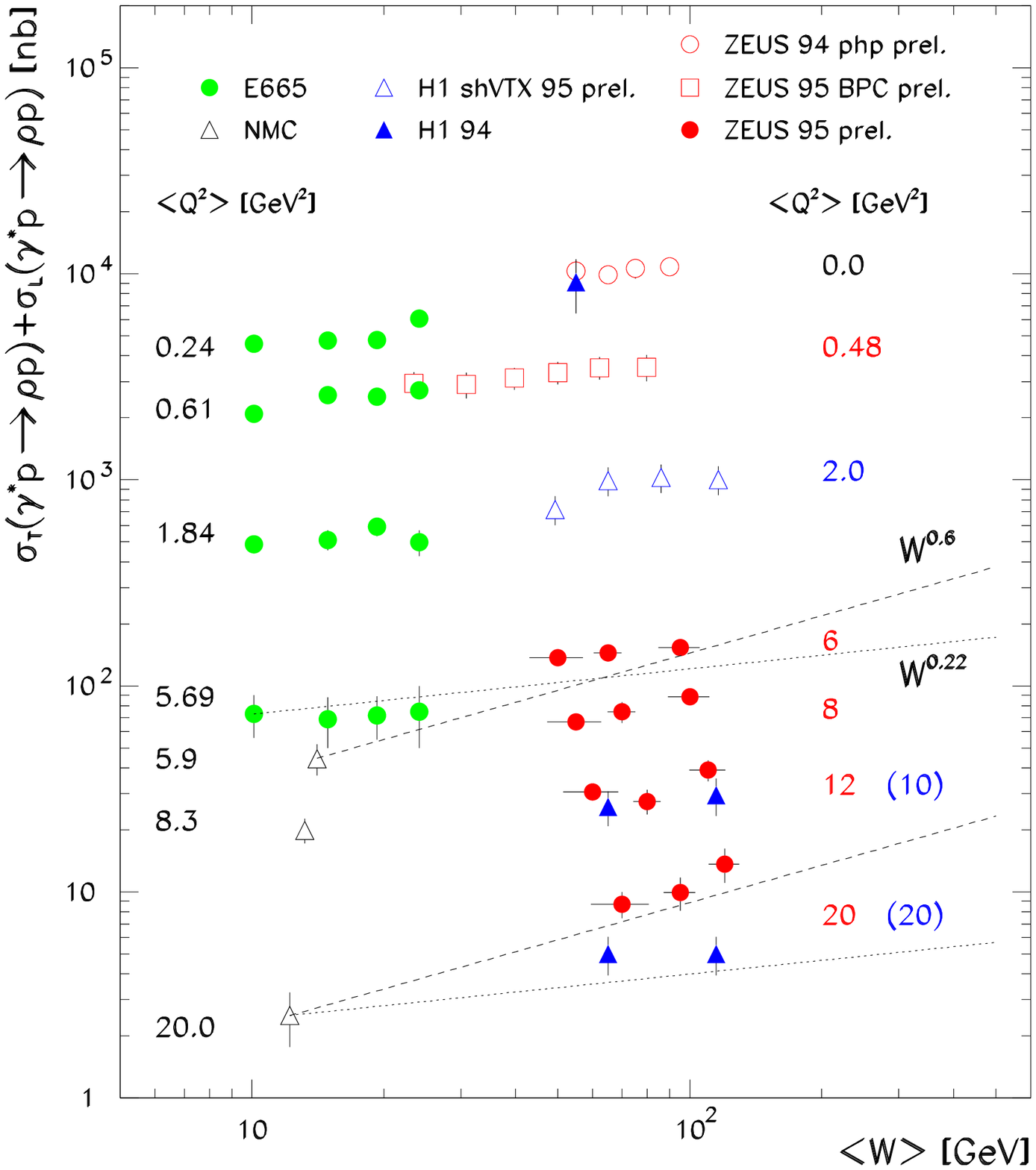}
\vspace*{-7mm}
\caption
{
\label{fig:disrho}
Sum of longitudinal and transverse elastic {\rhoz} electroproduction cross sections. The lines serve to indicate weak and strong power-law energy dependences. 
}
\end{minipage}
\hfill
\begin{minipage}[t]{0.48\textwidth}
\includegraphics[width=\textwidth, bb=0 22 496 534, clip=]{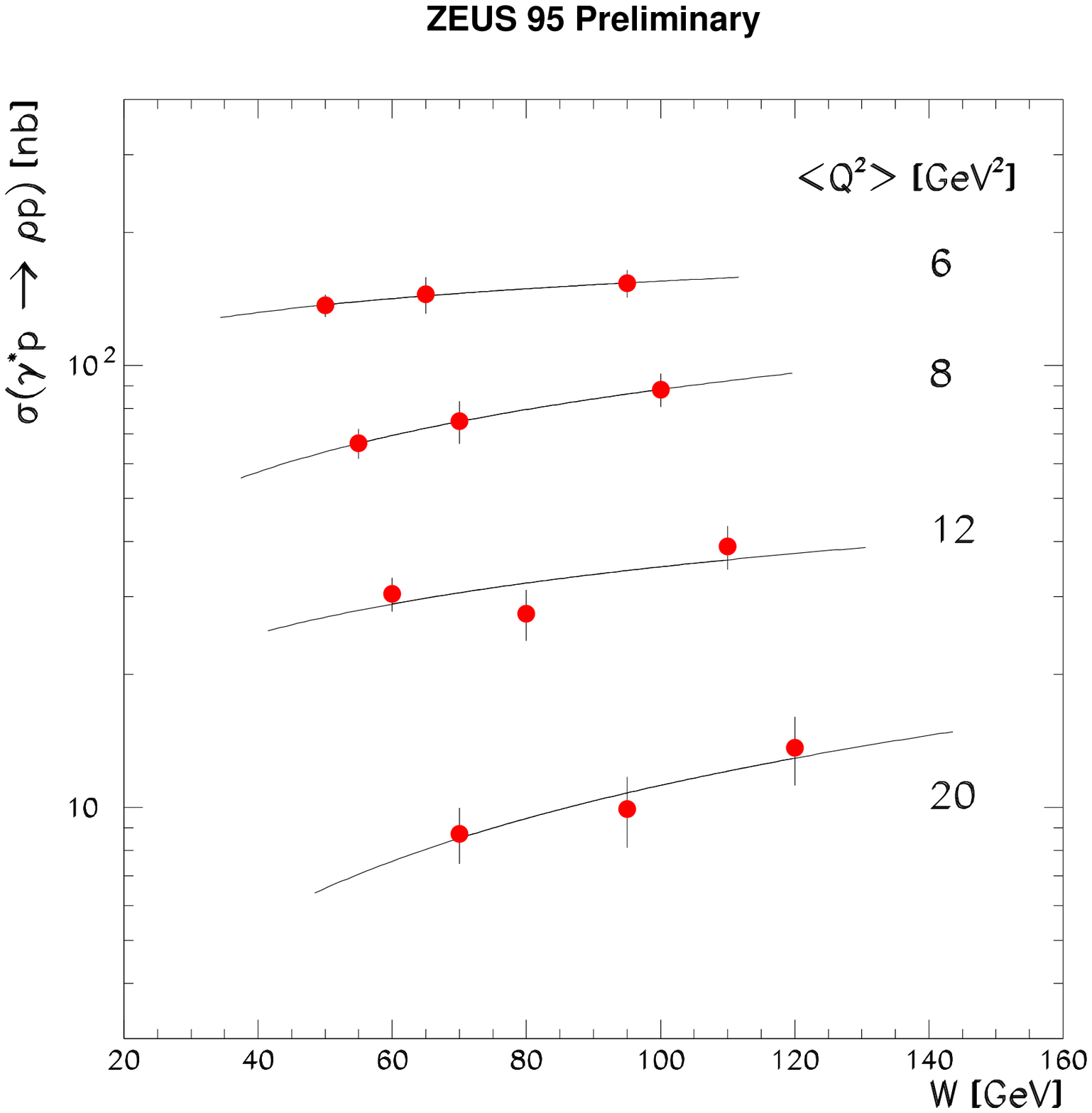}
\vspace*{-7mm}
\caption
{
\label{fig:disrhowdep}
Energy dependence of the elastic {\rhoz} electroproduction cross sections measured by the ZEUS collaboration at high {\qsq}~\protect\cite{j639}. 
}
\end{minipage}
%\vspace*{-3mm}
\end{figure}

The weak energy dependence of elastic {\rhoz} photoproduction was one of the prime indications that a process similar to that observed for the total hadronic cross sections is operative.  Here we address the question of the energy dependence for {\rhoz} production at high {\qsq}. Figure~\ref{fig:disrho} shows the present status of the elastic cross section measurements summed over transverse and longitudinal contributions, 
%\pagebreak
\begin{eqnarray*}
{\sigma}({\gptovp}) &=& \sigma_{\rm T}({\gptovp}) +  \sigma_{\rm L}(\gptovp) \\
 &=& \left( \frac{1+R}{1+{\varepsilon}R} \right) \; \left( \sigma_{\rm T}({\gptovp}) + {\varepsilon}\, \sigma_{\rm L}(\gptovp) \right)
\end{eqnarray*}
along with lines intended to indicate weak and strong types of energy dependence. The correction for the ratio of longitudinal to transverse photon flux, $\varepsilon$, was made using the determination of $R=\sigma_{\rm L}/\sigma_{\rm T}$ described in Sect.~\ref{sec:helicity}. (This correction is small compared to the accuracy of the measurement owing to the narrow range in $\varepsilon$ covered by the data: $0.97 < \varepsilon < 1.00$.) The statistical power of the high-{\qsq} measurements does not allow a sufficient determination of the energy dependence, and comparison to low-energy experiments is necessary. The measurements at {\qsq}=0.48~{\gevsq}, obtained with a special-purpose calorimeter installed in the ZEUS experiment near the beam-pipe in the electron flight direction (BPC), show a dependence similar to that for photoproduction. At higher {\qsq}, comparison to the low-energy experiments suffers from a discrepancy in the results from the NMC and E665 experiments, a discrepancy clearly evident in the E665 publication~\cite{zfp_72_237}. Furthermore, at the highest {\qsq} and $W$ points there is an inconsistency in the H1 and ZEUS results of approximately a factor of two. A conclusion thus awaits more work and/or measurements. Figure~\ref{fig:disrhowdep} 
exhibits an attempt by the ZEUS collaboration to extract an energy dependence from their data alone~\cite{j639}. A trend toward steeper energy dependence at higher {\qsq} is only weakly indicated by the data.
%\clearpage
\section[Momentum Transfer to the Proton as a Perturbative Scale]{Momentum Transfer to the Proton as a\\ Perturbative Scale}
\label{sec:tdep}
\subsection{Exponential Slopes}
\begin{figure}[htbp]
\vspace*{-2mm}
\begin{minipage}[t]{0.48\textwidth}
\includegraphics[width=\textwidth,bb= 0 10 530 546, clip=]{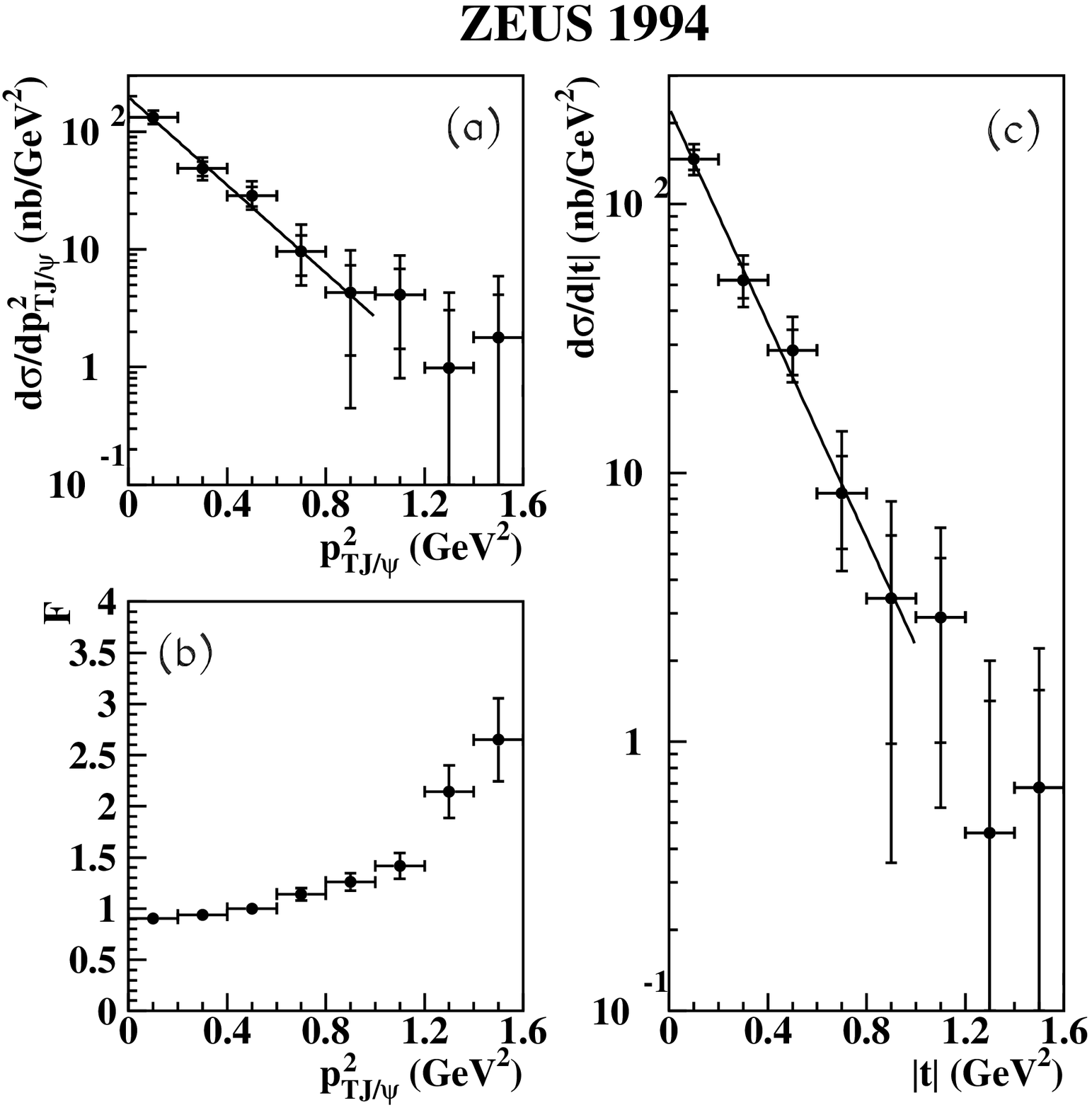}
\vspace*{-7mm}
\caption
{
\label{fig:pt2_t}
Elastic {\jpsi} photoproduction cross section: 1) the measured $p_{\mathrm T}^2$ dependence, b) the conversion factor $F$, calculated via simulation, used to convert the $p_{\mathrm T}^2$ dependence to the $|t|$ dependence,  and c) the result for the $|t|$ dependence, as
published by the ZEUS collaboration~\protect\cite{zfp_75_215}   \hspace*{\fill}
}
\end{minipage}
\hfill
\begin{minipage}[t]{0.48\textwidth}
\includegraphics[width=\textwidth, bb= 19 142 523 648, clip=]{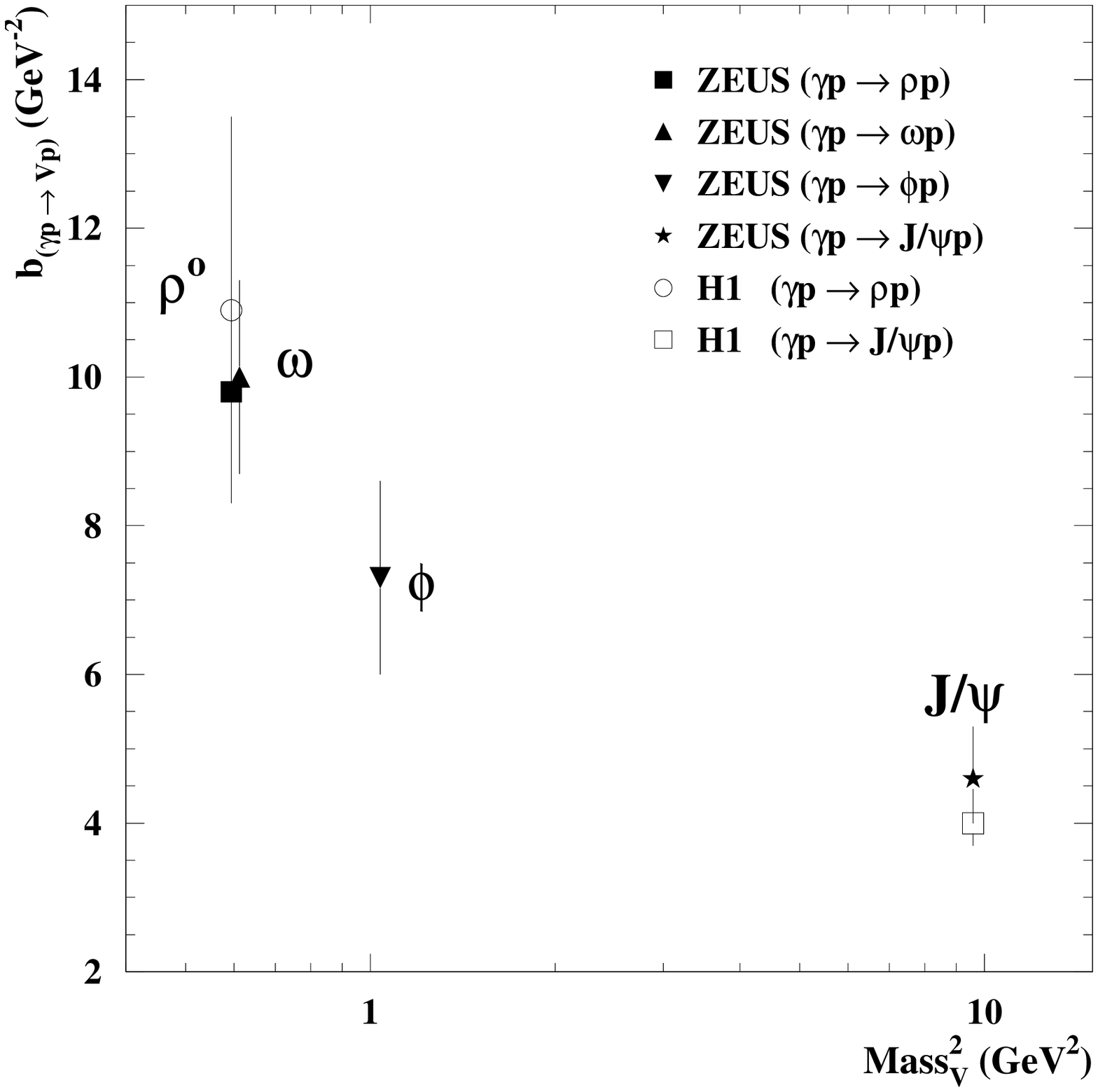}
\vspace*{-7mm}
\caption
{
\label{fig:vmpho_b_mass}
Exponential slope in the $t$-dependence of the photoproduction cross section plotted as a function of mass for each of the vector mesons investigated by the H1 and ZEUS collaborations
}
\end{minipage}
\vspace*{-2mm}
\end{figure}

As described in Sect.~\ref{sec:signatures}, the exponential $t$-slope in elastic scattering yields information on the spatial extent of the interaction, which has contributions from the size of the proton and the size of the vector meson.
The momentum transfer $t$ is closely related to the negative squared transverse momentum of the elastically produced vector meson in the laboratory frame; the two quantities differ by less than the value of {\qsq}. Thus the $t$-slope is accessible even when neither the final-state proton nor the final-state electron is detected, provided {\qsq} is limited to less than a few~{\gevsq}, a condition satisfied by requiring the electron to have escaped the central detector via the beam pipe in the electron flight direction.
Figure~\ref{fig:pt2_t} shows the extraction of the $|t|$ dependence
by the ZEUS collaboration in elastic {\jpsi} photoproduction~\cite{zfp_75_215}:
({a}) the dependence of the  {\jpsi} photoproduction
cross section on the squared transverse momentum of the
{\jpsi}, $p_{\mathrm T}^2$, ({b}) the factor, $F$, required
to relate the  $p_{\mathrm T}^2$ dependence
to the $|t|$ dependence, derived 
from simulations,
({c}) the $|t|$-dependence of the {\jpsi} elastic photoproduction cross
section. The result for the exponential slope, $4.6 \pm 0.4 \;{\rm (stat)} ^{+0.11}_{-0.16} \;{\rm (sys)}\; {\gevsq}$, is compared to values obtained for the {\rhoz}, $\omega$, and $\phi$ mesons in Fig.~\ref{fig:vmpho_b_mass}.
The trend toward smaller {\qqbar} configurations at higher vector-meson mass is clearly observed. The value for the {\jpsi} is consistent with the value of about 5 {\gev} determined for the proton alone from pp elastic and single dissociative scattering measurements, from which one can conclude that the {\jpsi} is smaller than the proton.

\begin{figure}[htbp]
\vspace*{-3mm}
\begin{minipage}[b]{0.48\textwidth}
\hspace*{3mm}The large {\qsq} range covered by the elastic {\rhoz} leptoproduction measurements at HERA is shown in Fig.~\ref{fig:vm_b_q2_new}, 
together with E665 results at low energy. A trend toward decreasing size of the {\qqbar} configuration with {\qsq}, as predicted within the framework of QCD, is confirmed. \vspace*{1cm}
\caption
{
\label{fig:vm_b_q2_new}
The exponential slopes of the elastic {\rhoz} production cross sections as a function of photon virtuality {\qsq}
}
\end{minipage}
\hfill
\begin{minipage}[t]{0.48\textwidth}
\includegraphics[width=\textwidth, bb= 0 13 520 515, clip=]{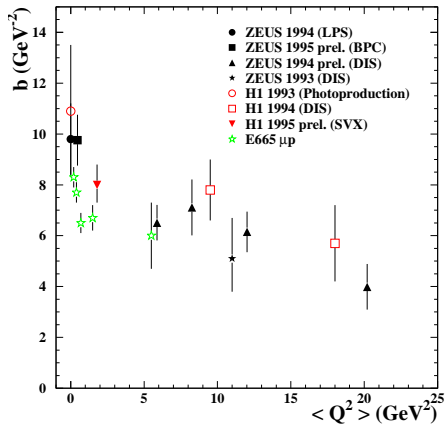}
\end{minipage}
\vspace*{-2mm}
\end{figure}

%\clearpage
\subsection{Shrinkage}
A question important to the determination of the fundamental physical process underlying diffractive vector-meson production is whether a value for $\alpha_{\Pma}'$ (see Sect.~\ref{sec:signatures}) can be measured, and whether the result is consistent with that derived from total hadronic cross sections. 
\begin{figure}[htbp]
\vspace*{-2mm}
\begin{minipage}[b]{0.44\textwidth}
Figure~\ref{fig:bvswrho} shows a comparison by the ZEUS collaboration~\cite{epj_2_247} of their measurements for the exponential slope $b$ with those at low energy in order to 
test consistency with Eq.~\ref{eq:b0}. Despite the long lever arm in energy 
provided by the HERA data, the measurements do not allow an unambiguous conclusion. A fit to the ZEUS data points alone yields the result $\alpha_{\Pma}'\, =\, 0.23 \pm 0.15 \;{\rm (stat)} ^{+0.10}_{-0.07} \;{\rm (sys)}\; {\gev}^{-2}$~\cite{dirk}, which is shown as the line in Fig.~\ref{fig:bvswrho}.
\end{minipage}
\hfill
\begin{minipage}[b]{0.52\textwidth}
\includegraphics[width=\textwidth, bb= 10 260 455 516, clip=]{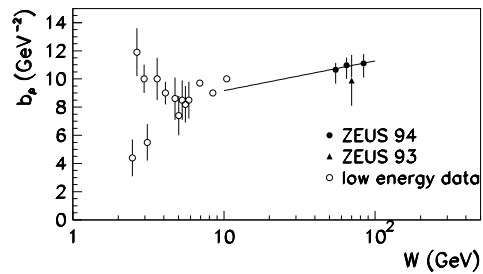}
\vskip -3mm
\caption
{
\label{fig:bvswrho}
The exponential slope of the $t$ distribution for elastic {\rhoz} photoproduction as a function of the center-of-mass energy $W$. The line indicates the fit result described in the text~\protect\cite{dirk}
}
\end{minipage}
\vspace*{-2mm}
\end{figure}

\begin{figure}[htbp]
\vspace*{-2mm}
\begin{minipage}[t]{0.6\textwidth}
\includegraphics[width=\textwidth, bb= 0 17 562 545, clip=]{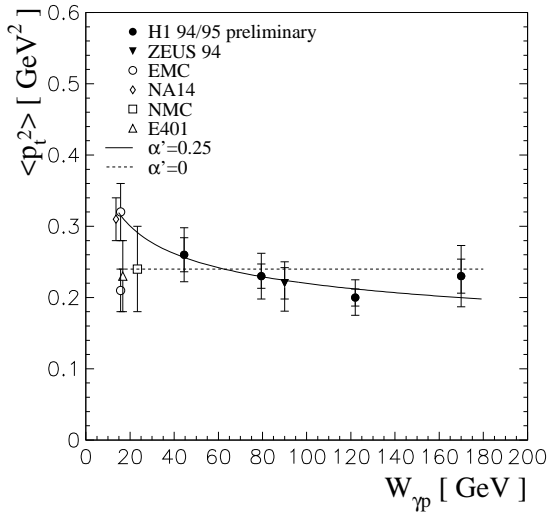}
\end{minipage}
\hfill
\begin{minipage}[b]{0.37\textwidth}
\caption
{
\label{fig:psishr_h1}
The average transverse momentum of elastically photoproduced {\jpsi} mesons as a function of the center-of-mass energy $W$
}
\end{minipage}
\vspace*{-2mm}
\end{figure}

The H1 collaboration has presented an analysis of the average transverse momentum of {\jpsi} mesons in elastic photoproduction~\cite{j242, *west}, comparing the result to ZEUS results and to measurements at low energy. Figure~\ref{fig:psishr_h1} shows the results of the analysis, as well as the expected behavior for $\alpha_{\Pma}'=0.25\;{\gev}^{-2}$ and for no $t$-dependence at all. The data are consistent with each of the two assumptions.

Last December, A.~Levy put out a preprint describing an analysis~\cite{hep97_12_519} in which he attempted to determine $\alpha_{\Pma}$ directly by fitting the exponent in an assumed power-law $W$-dependence for the value of $\D\sigma/\D{t}$ in five distinct $t$-bins, employing ZEUS, H1, and low-energy data from the EMC and E401 experiments.{\kern-4.pt}~\footnotemark 
\begin{figure}[htbp]
\vspace*{-2mm}
\begin{minipage}[b]{0.48\textwidth}
The results are shown in Fig.~\ref{fig:psishr},
and exclude with a high degree of confidence the shrinkage exhibited by the total hadronic cross sections. Along with the steep energy dependence of the elastic {\jpsi} photoproduction cross section, this absence of shrinkage points to the validity of a pQCD approach in describing this process.\vspace*{5mm}
\caption
{
\label{fig:psishr}
The value for $\alpha_{\Pma}$ as a function of $t$ determined from fits to
the $W$-dependence for the differential elastic {\jpsi} photoproduction cross sections as described in the text~\protect\cite{hep97_12_519}. The result excludes a shrinkage effect of the magnitude observed in total hadronic cross sections
}
\end{minipage}
\hfill
\begin{minipage}[b]{0.46\textwidth}
\includegraphics[width=\textwidth, bb= 24 68 526 767, clip=]{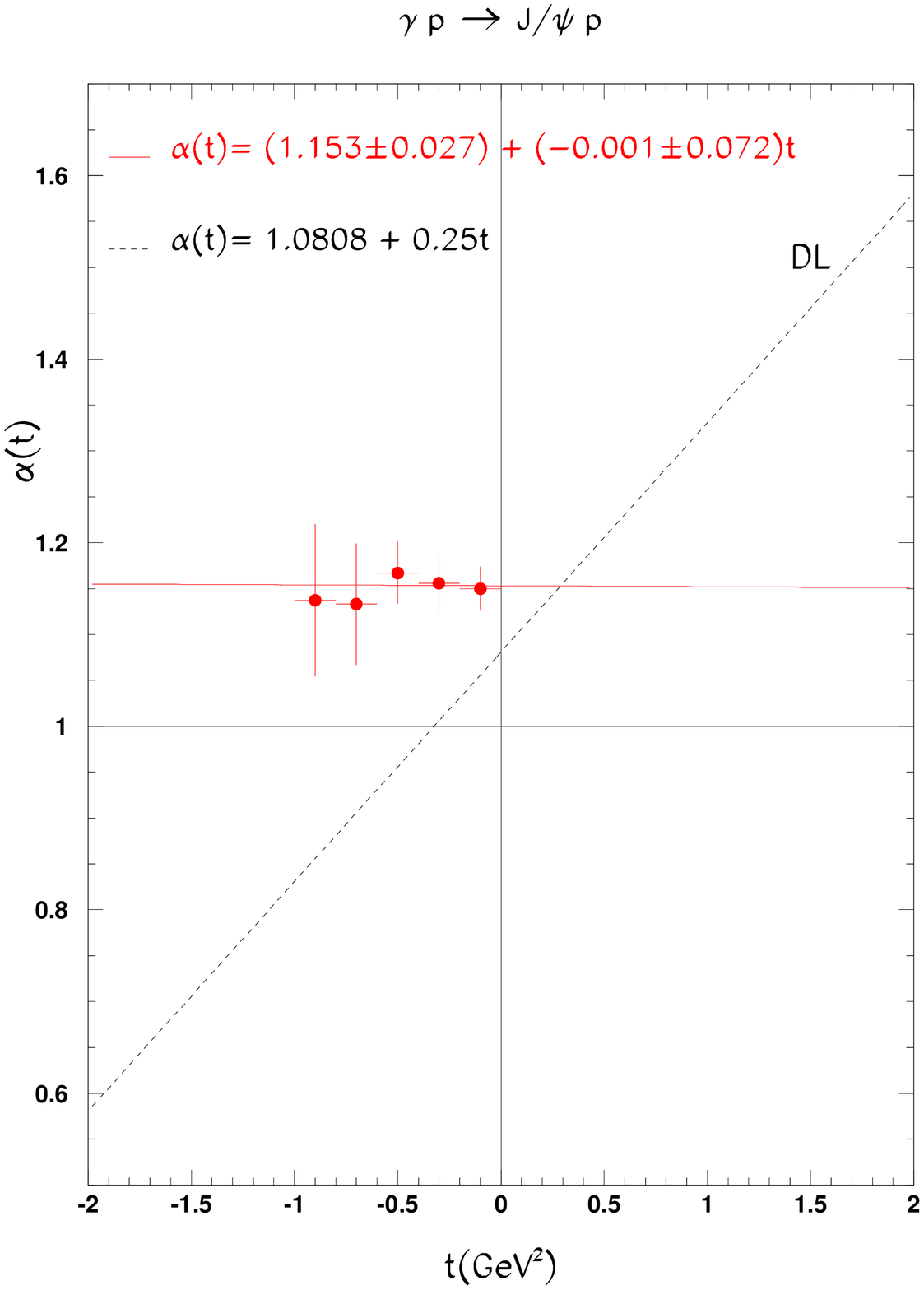}
\end{minipage}
\vspace*{-2mm}
\end{figure}
\footnotetext{A major distinction between Levy's approach and that of the H1 analysis is Levy's insensitivity to the factor 
$e^{b_0t}$, since he restricts his consideration to the $W$ dependence. The quantity $<p_T^2>$ employed in the H1 analysis is primarily determined by the $e^{b_0t}$ factor (see Eq.~\ref{eq:ebt}.)}

%\clearpage
\subsection{Dipion Mass Spectrum in {\rhoz} Photoproduction}
\begin{figure}[htbp]
\vspace*{-2mm}
\begin{minipage}[t]{0.48\textwidth}
\includegraphics[width=\textwidth,bb=0 19 462 485,clip=]{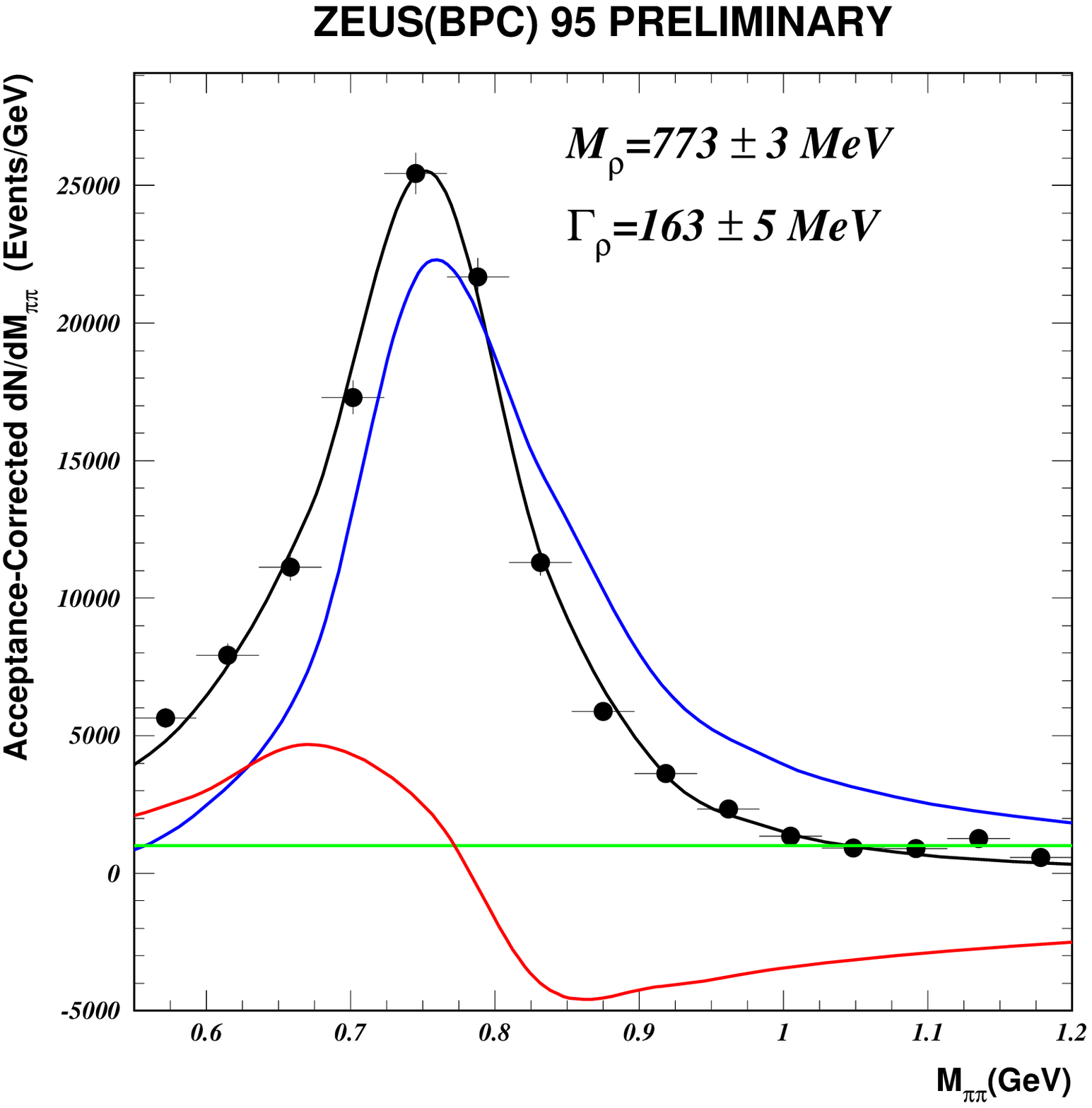}
\vspace*{-7mm}
\caption{
\label{fig:bpc_mass_new}
Acceptance-corrected $\pi^+\pi^-$ invariant mass distribution
for elastic {\rhoz} production for the region $0.25 < {\qsq} < 0.85$. The curves are the result of a fit to the function in Eq.~\protect\ref{eq:dipionspectrum}
\hspace*{\fill}
}
\end{minipage}
\hfill
\begin{minipage}[t]{0.48\textwidth}
\includegraphics[width=\textwidth]{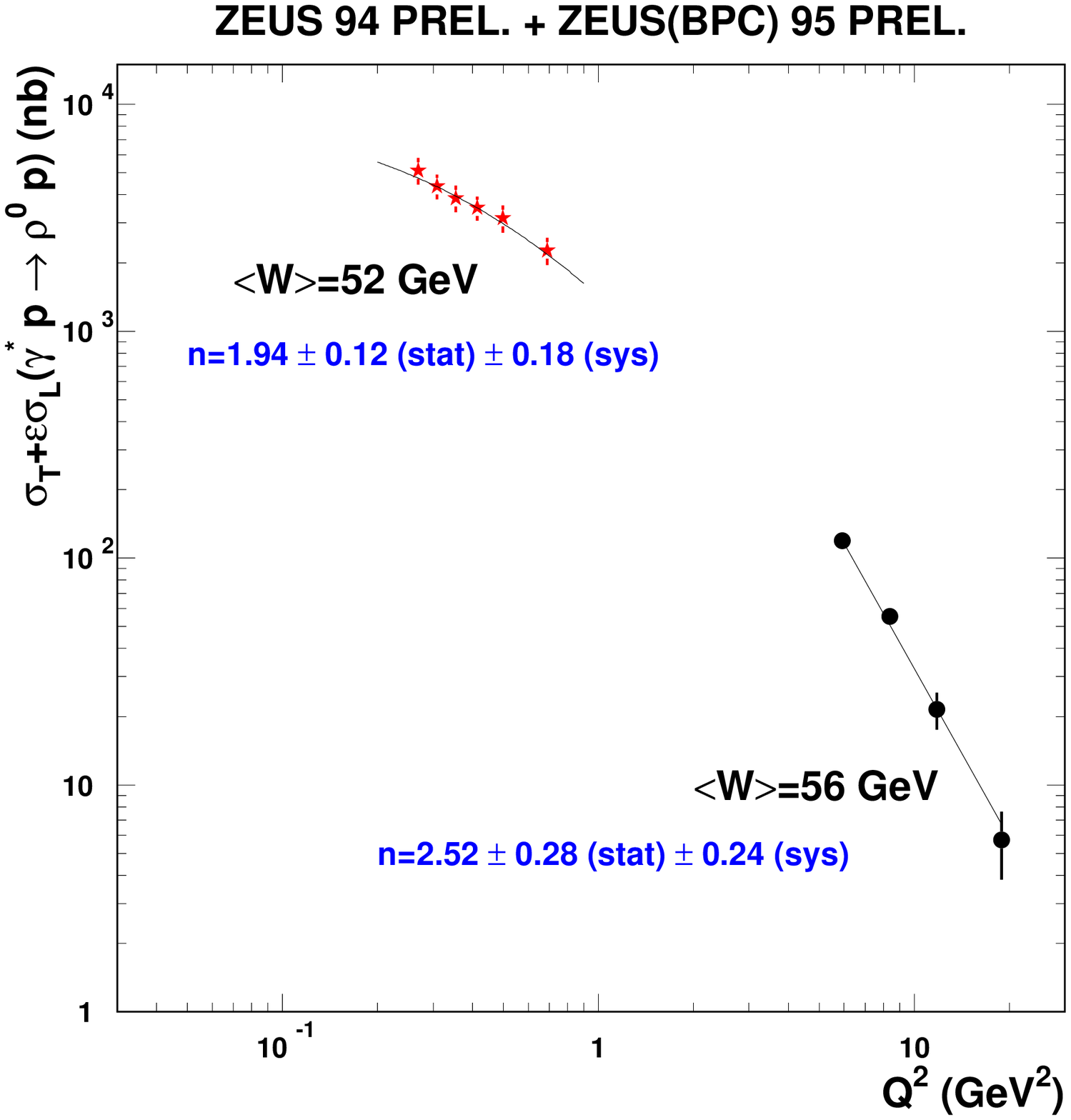}
\vspace*{-7mm}
\caption{
\label{fig:bpc_q2plot}
The {\qsq} dependence of the {\rhoz} elastic cross section. See text for details  \hspace*{\fill}
}
\end{minipage}
\vspace*{-2mm}
\end{figure}

The $\pi^+\pi^-$ mass spectrum in low-$|t|$ {\rhoz} photoproduction exhibits an asymmetry which can be described by the interference with continuum dipion production, as described in Sect.~\ref{sec:signatures}:
\begin{eqnarray}
\label{eq:dipionspectrum}
\frac{dN}{dM_{\pi \pi}} &=& \left | {A} \frac{\sqrt{M_{\pi \pi}M_{\rho}\Gamma_{\rho}}}{M^2_{\pi \pi}-M^2_{\rho}+iM_{\rho}\Gamma_{\rho}}+ {B} \right |^2\;.
\end{eqnarray}
 The ZEUS collaboration has employed a data set based on their beam-pipe calorimeter to quantify this skewing effect in the region $0.25 < {\qsq} < 0.85$~\cite{teresa,hep97_09_031}, as shown in Fig.~\ref{fig:bpc_mass_new}. Extracting the elastic {\rhoz} cross section from the result, they studied the {\qsq} dependence by comparing the cross section measured at ${\qsq} \approx M_{\rho}^2$ with the results at high {\qsq} and found the data to be consistent with the same value of the exponent $n$ when using the parametrization $({\qsq}+M_{\rho}^2)^{-n}$, as shown in Fig.~\ref{fig:bpc_q2plot}.

\begin{figure}[htbp]
\vspace*{-2mm}
\begin{minipage}[t]{0.60\textwidth}
\raisebox{-1mm}{\includegraphics[width=\textwidth, bb= 0 0 456 487, clip=]{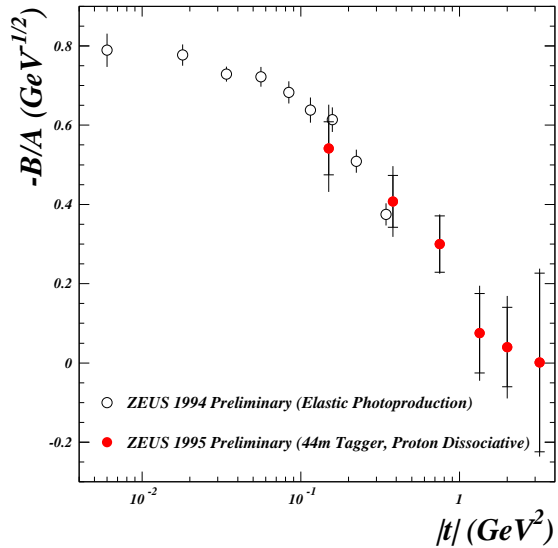}}
\end{minipage}
\hfill
\begin{minipage}[b]{0.36\textwidth}
\caption
{
\label{fig:ba_44m}
Ratio of the $B$ to $A$ parameter (see Eq.~\protect\ref{eq:dipionspectrum}) in diffractive $\pi^+\pi^-$ photoproduction in the {\rhoz} mass region as presented by the ZEUS collaboration~\protect\cite{epj_2_247,dirk,hep97_09_031,leszek}
}
\end{minipage}
\vspace*{-2mm}
\end{figure}

The ZEUS collaboration installed another small electron calorimeter near the beam pipe 44 m from the interaction point prior to the 1995 running period, with acceptance limited to ${\qsq}<0.01\,\gevsq$. In this manner they isolate a sample of high $|t|$ {\rhoz} mesons, identifying the transverse momentum of the {\rhoz} with the momentum transfer to the undetected proton. This extends the study of the mass spectrum skewing to higher $|t|$ than was possible with the high-statistics elastic photoproduction investigation without the photon tag~\cite{epj_2_247}, as shown in Fig.~\ref{fig:ba_44m}. The results clearly show the relative contribution of the nonperturbative continuum to decrease with $|t|$, and to become consistent with zero for \mbox{$|t|>1\;\gevsq$.}
%\clearpage
\subsection{Vector-Meson Production Ratios}
\begin{figure}[htbp]
\vspace*{-2mm}
\begin{minipage}[b]{0.41\textwidth}
The same tagged photoproduction sample was used by the ZEUS collaboration to study the production of $\phi$ and {\jpsi} mesons relative to that for the {\rhoz} as a function of $t$. Figure~\ref{fig:ratios_44m} shows the ratios to be rising toward the values of 2/9 and 8/9 expected from a flavor-independent point-like electromagnetic coupling to the valence quark content of the vector mesons.
% \vspace*{5mm}
\caption
{
\label{fig:ratios_44m}
Ratios of $\phi$ and {\jpsi} production to {\rhoz} photoproduction
as a function of $t$. See text for details  \hspace*{\fill}
}
\end{minipage}
\hfill
\begin{minipage}[b]{0.55\textwidth}
\includegraphics[width=\textwidth, bb= 0 7 514 543, clip=]{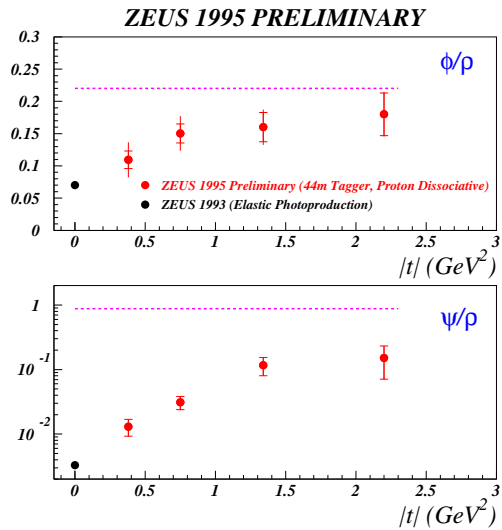}
\end{minipage}
\vspace*{-2mm}
\end{figure}

\subsection{{\jpsi} Photoproduction}
\begin{figure}[htbp]
\vspace*{-2mm}
\begin{minipage}[t]{0.48\textwidth}
\includegraphics[width=\textwidth, bb= 0 245 514 745, clip=]{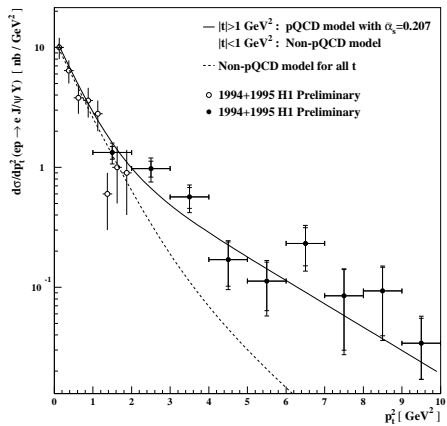}
\end{minipage}
\hfill
\begin{minipage}[b]{0.48\textwidth}
\caption
{
\label{fig:psi_hight_h1}
The {\jpsi} photoproduction cross section as a function of the squared transverse momentum of the {\jpsi} as reported by the H1 collaboration~\protect\cite{j274}. The solid line is the result of the pQCD calculation described in ref.~\protect\cite{pl_375_301}
}
\end{minipage}
\vspace*{-2mm}
\end{figure}

The H1 collaboration has studied {\jpsi} photoproduction at high values for the {\jpsi} transverse momentum, as shown in Fig.~\ref{fig:psi_hight_h1}. 
They find that a good description of the measurements is given by the pQCD-based calculation of ref.~\cite{pl_375_301}. \newpage

%\clearpage
\section{Helicity Analyses}
\label{sec:helicity}
A firm prediction of QCD for the diffractive leptoproduction of vector mesons is that the cross section for longitudinally polarized photons will exceed that for transversely polarized photons at high {\qsq}~\cite{pr_50_3134, pr_56_2982}. The present status of the HERA analyses of the vector meson decay angular distributions allow the extraction of $R=\sigma_{\rm L}/\sigma_{\rm T}$ only under the assumption that helicity is conserved in the $s$-channel amplitudes for the $\gamma^*/{\rhoz}$ transition (SCHC). While this assumption is justified by the results of low-energy experiments, and is supported by consistency checks using the HERA data, the statistical power of the data has not yet permitted a complete determination of the spin-density matrix elements. The purpose of this section is to introduce the mechanics of the helicity analyses and present the current status of determinations of $R$ as a function of {\qsq}.

\subsection[Decay Angle Distributions and $s$-Channel Helicity Conservation]{Decay Angle Distributions\\ and $s$-Channel Helicity Conservation}
The diffractive dissociation of a real photon into a massive vector meson presents two fundamental questions: 1) How do the two polarization degrees of freedom
of the photon relate to the three degrees of freedom available to the vector meson? 2) We expect helicity to be conserved in this strong interaction, but in what frame? A priori it is unclear whether spin should be quantized along the photon direction or along the vector meson direction. Schilling et al.~\cite{np_15_397} addressed the first question by decomposing the {\rhoz} decay angular distribution into contributions from the various polarization states of the photon:
\begin{itemize}
\item
unpolarized: $W(\thetah,\phih)=W^0(\thetah,\phih)$,
\item
circularly polarized: $W(\thetah,\phih)=W^0(\thetah,\phih)\pm P^c_{\gamma}\,W^3(\thetah,\phih)$,
\item
%linearly polarized:$W(\thetah,\phih)=W^0(\thetah,\phih) - P^l_{\gamma}\,\cos{2\Phih}\,W^3(\thetah,\phih)- P^l_{\gamma}\,\sin{2\Phih}\,W^3(\thetah,\phih)$,
linearly polarized:  $W(\thetah,\phih,\Phih) = {W^0}(\thetah,\phih)$ 
\parbox[t]{8cm}{
-  $P^l_{\gamma} \; \cos{2\Phih} \;{W^1}(\thetah,\phih)$\\
-  $P^l_{\gamma} \; \sin{2\Phih} \;{W^2}(\thetah,\phih)$,
}

\end{itemize}
where $P^c_{\gamma}$ and $P^l_{\gamma}$ are the degrees of circular and linear polarization. The angles are defined with respect to the planes shown in Fig.~\ref{fig:heldefs} as follows:
\begin{figure}[htbp]
\vspace*{-2mm}
\begin{minipage}[t]{0.48\textwidth}
\includegraphics[width=\textwidth, bb= 109 112 801 630, clip=]{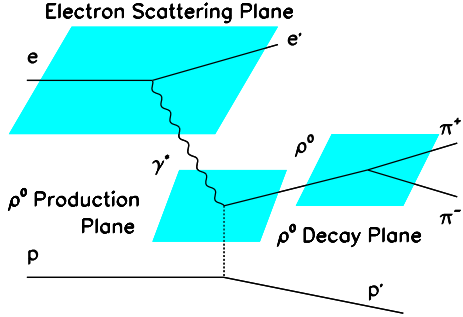}
\end{minipage}
\hfill
\begin{minipage}[b]{0.48\textwidth}
\caption
{
\label{fig:heldefs}
Schematic diagram of the planes in diffractive {\rhoz} production used to define the helicity angles as specified in the text \hspace{\fill}
}
\end{minipage}
\vspace*{-2mm}
\end{figure}
{\thetah} and {\phih} are the polar and azimuthal angles of the
momentum of the
positively charged vector-meson decay product in the coordinate system where
the $z$ axis is defined to be along the
the vector-meson momentum
in the $\gamma{\rm p}$ center-of-mass frame and
the $y$ axis is perpendicular to
the vector-meson production
plane, and where {\Phih} is the
angle between the vector-meson production plane and the electron
scattering plane. They derived expressions for these angular distributions in terms of the spin-density matrix elements describing the vector-meson spin state, the $W^\alpha$ depending on the elements $\rho^{\alpha}_{ik}$, with $i,k=-1,0,+1$ representing the three polarization states of the vector meson. Thus the element $\rho^0_{00}$ represents the probability for producing a longitudinally polarized vector meson from unpolarized transverse photons, $\rho^1_{1-1}$ corresponds to the interference between the amplitudes for producing vector mesons in the helicity states 1 and -1 from linearly polarized photons in the helicity state 1, and so on.

\begin{figure}[htbp]
\vspace*{-2mm}
\begin{minipage}[t]{0.48\textwidth}
\includegraphics[width=\textwidth]{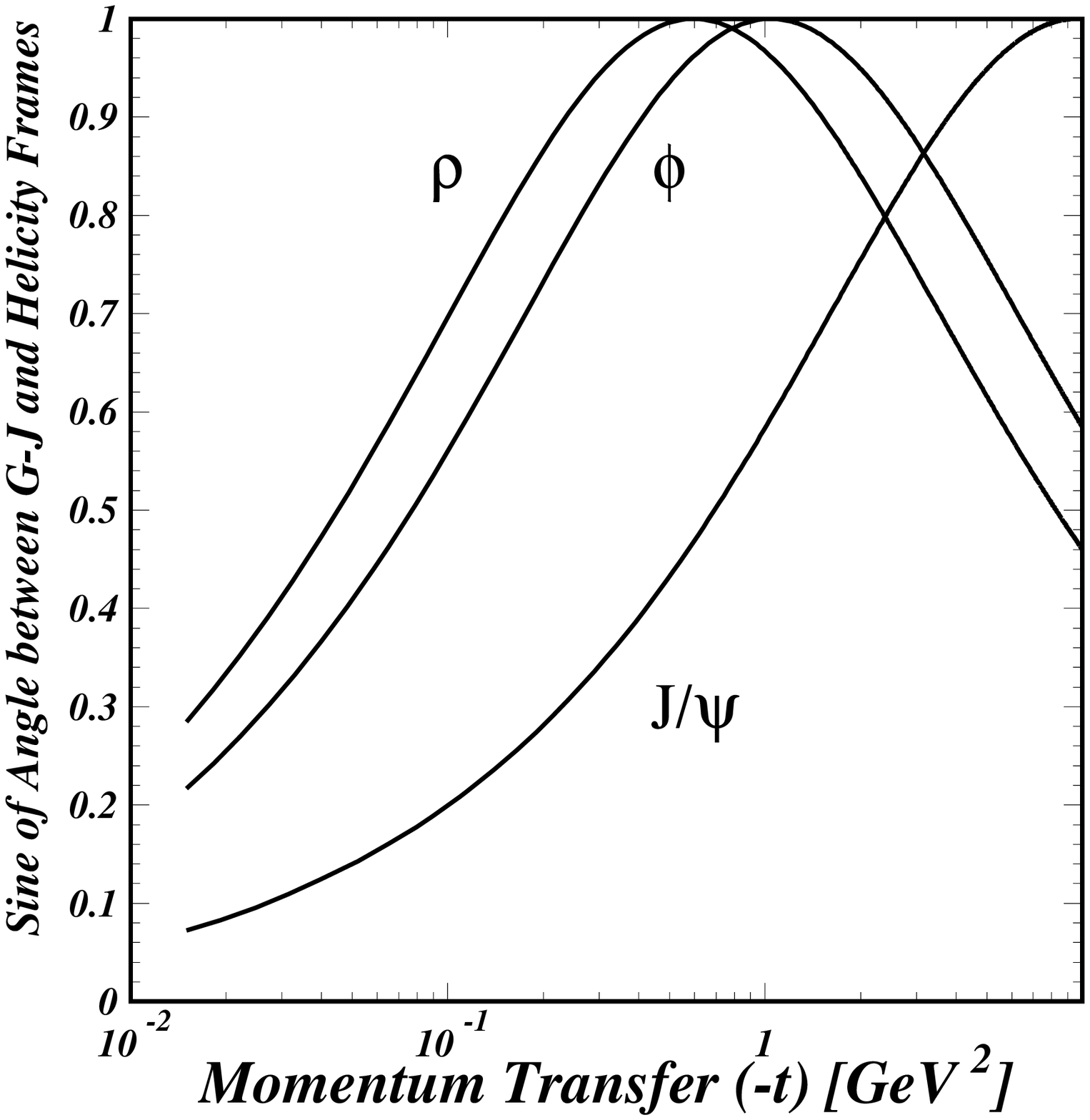}
\vspace*{-7mm}
\caption
{
\label{fig:gj}
The sine of the angle between the Gottfried-Jackson and $s$-channel helicity frames as a function of $t$ for {\rhoz}, $\phi$, and {\jpsi} mesons} \hspace{\fill}
\end{minipage}
\hfill
\begin{minipage}[t]{0.48\textwidth}
\includegraphics[width=\textwidth, bb=34 9 513 561, clip=]{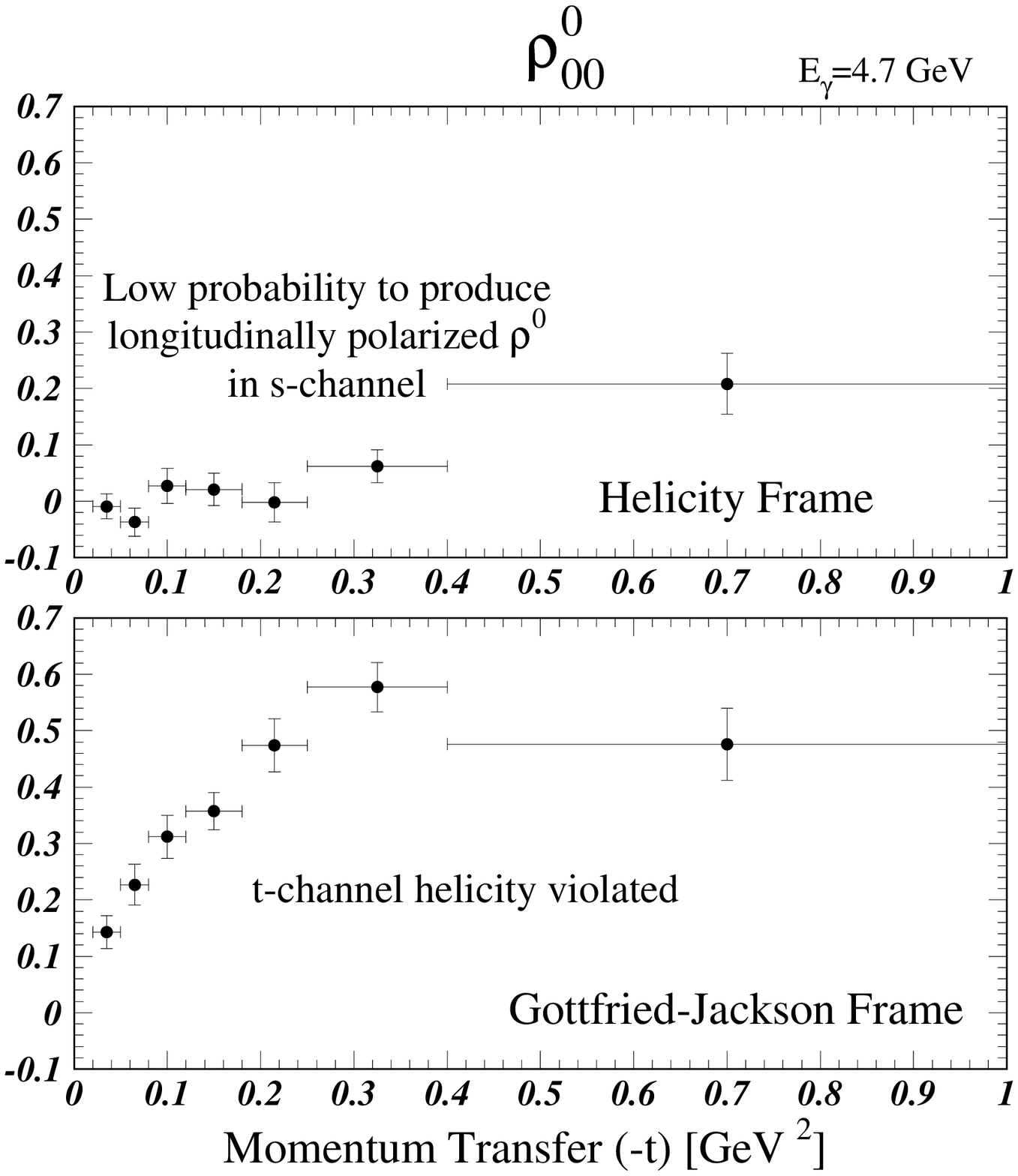}
\vspace*{-7mm}
\caption
{
\label{fig:ballam}
The spin density matrix element $\rho^0_{00}$ as a function of $t$ measured by Ballam et al.~\protect\cite{pr_5_545} \hspace{\fill}
}
\end{minipage}
\vspace*{-2mm}
\end{figure}

The second question was addressed in detail by Ballam et al. in the early 1970s, when they studied {\rhoz} photoproduction in a bubble chamber using a
linearly polarized back-scattered laser beam. They compared the matrix elements calculated from the decay angular distributions in a coordinate system where the spin quantization axis is aligned with the photon direction (the Gottfried-Jackson frame) with those calculated in the $s$-channel helicity frame, where the 
spin is quantized along the vector-meson direction in the $\gamma p$ center-of-mass frame. Figure~\ref{fig:gj} shows the sine of the angle between those frames as a function of $t$ for {\rhoz}, $\phi$, and {\jpsi} mesons. It is particularly interesting to note that for {\jpsi} photoproduction significantly higher values of $|t|$ must be reached before the discriminating power between the two hypotheses becomes comparable to that in the case of the {\rhoz} or $\phi$ meson. Figure~\ref{fig:ballam} shows the measurements by Ballam et al.~\cite{pr_5_545} which confirmed $s$-channel helicity conservation (SCHC). The matrix element $\rho_{00}^0$, which represents the probability to produce longitudinally polarized {\rhoz} mesons from the transverse photons, is consistent with zero for $|t|<0.5\;\gev$. This measurement prompted Gilman et al. to point out that the result is inconsistent with the hypothesis of $t$-channel exchange of a spinless particle~\cite{pl_31_387}.

Schilling and Wolf~\cite{np_61_381} generalized this treatment to the case of virtual photons ($R>0$). They noted that the decay angular distributions for unpolarized transverse photons are indistinguishable from those for longitudinally polarized photons, and therefore defined the following linear combinations of the matrix elements:
\begin{eqnarray*}
r^{04}_{ik} &=& {\rho^0_{ik} \, + \, \varepsilon R \rho^{4}_{ik}\over 1 \, + \, \varepsilon R} \\[5mm]
r^{\alpha}_{ik} &=& \,
\left\{
\begin{array}{ll}
{\displaystyle {\rho^{\alpha}_{ik}\over 1 \, + \, \varepsilon R}}, & {\alpha}=1-3 \hspace*{4mm}{\rm (transverse)}\\*[5mm]
{\displaystyle {\sqrt{R} \; \rho^{\alpha}_{ik}\over 1 \, + \, \varepsilon R}}, & {\alpha}=5-8 \hspace*{4mm}{\rm (long./trans.\hspace*{1.5mm} interference)}.
\end{array} \right.
\end{eqnarray*}
%
%\begin{eqnarray*}
%r^{04}_{ik} &=& \frac{\rho^0_{ik} \, + \, \epsilon { R} \rho^{4}_{ik}}{1 \, + \, \epsilon { R}}\\[5mm]
%r^{\alpha}_{ik} &=& \, \left\{ \begin{array}{ll} {{\rho^{\alpha}_{ik}}\over{1 \, + \, \epsilon \strut { R}}} & {\alpha}=1-3\;{\mbox{\small (transverse)}}\\[5mm] {\sqrt{{ R}} \; \rho^{\alpha}_{ik}}\over{1 \, + \, \epsilon \strut { R}} & {\alpha}=5-8\;{\mbox{\small (long./trans. interference).}} \end{array} \right.
%\end{eqnarray*}

The $R$ dependence of these terms identifies them as arising from transverse or longitudinal amplitudes, or from the interference of the two. Circular photon polarization ($\alpha = 3$) in leptoproduction requires longitudinally polarized initial-state leptons, so the H1 and ZEUS experiments will not have information on $r^3_{ik}$ until spin rotators are installed, which is a few years off. However, the E665 experiment has published results~\cite{zfp_72_237} for circularly polarized photons (the initial-state muons are produced longitudinally polarized via pion decay), and the HERMES experiment~\cite{dueren} is also sensitive to these matrix elements due to their spin rotators.

\subsection{Measurements of the {\qsq} Dependence of $R$}
The assumption of SCHC allows the ratio $R$ to be determined from the distribution in the polar angle {\thetah} of the $\pi^+$ in the {\rhoz} rest frame by measuring the probability $r_{00}^{04}$ that the {\rhoz} is produced  longitudinally polarized:
\begin{eqnarray*}
\frac{1}{N} \frac{\D N}{\D (\cos{\thetah})}
&=&\frac{3}{4}\left[1-r_{00}^{04}+(3r_{00}^{04}-1)\cos^2{\thetah}\right],
\end{eqnarray*}
and deriving $R$ via $R = {r_{00}^{04}}/ \left[\varepsilon (1-r_{00}^{04})\right]$.

\begin{figure}[htbp]
\vspace*{-2mm}
\begin{minipage}[t]{0.48\textwidth}
\begin{center}
\includegraphics[width=0.8\textwidth, bb= 6 57 513 544, clip=]{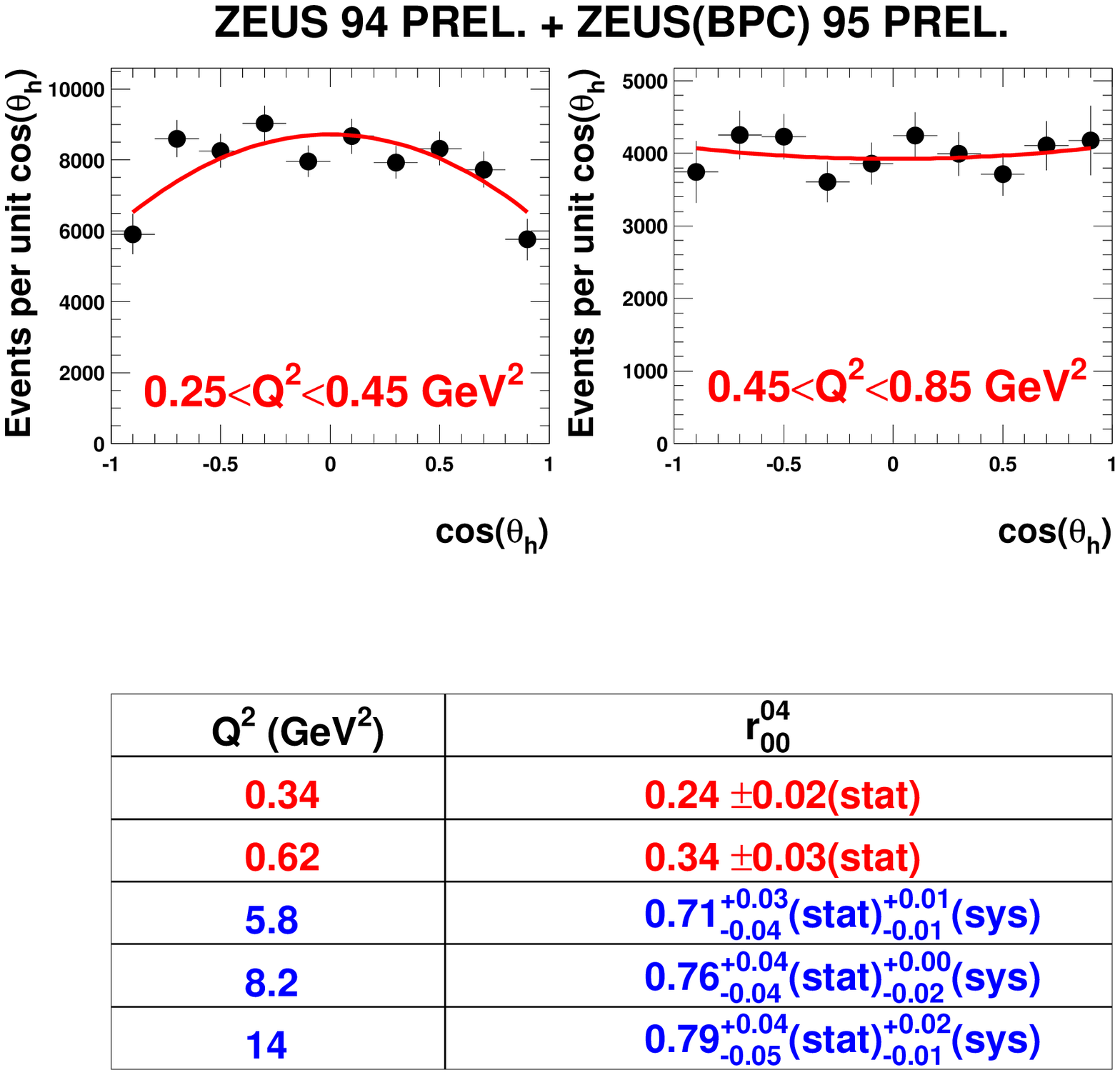}
\vspace*{-3mm}
\caption
{
\label{fig:bpc_hel_theta}
Polar decay angle distributions for the ZEUS {\rhoz} samples and
fit results for $r_{00}^{04}$ 
at intermediate and high {\qsq}~\protect\cite{teresa, hep97_09_031}  \hspace*{\fill}
}
\end{center}
\end{minipage}
\hfill
\begin{minipage}[t]{0.48\textwidth}
\begin{center}
\includegraphics[width=0.8\textwidth, bb= 10 16 456 463, clip=]{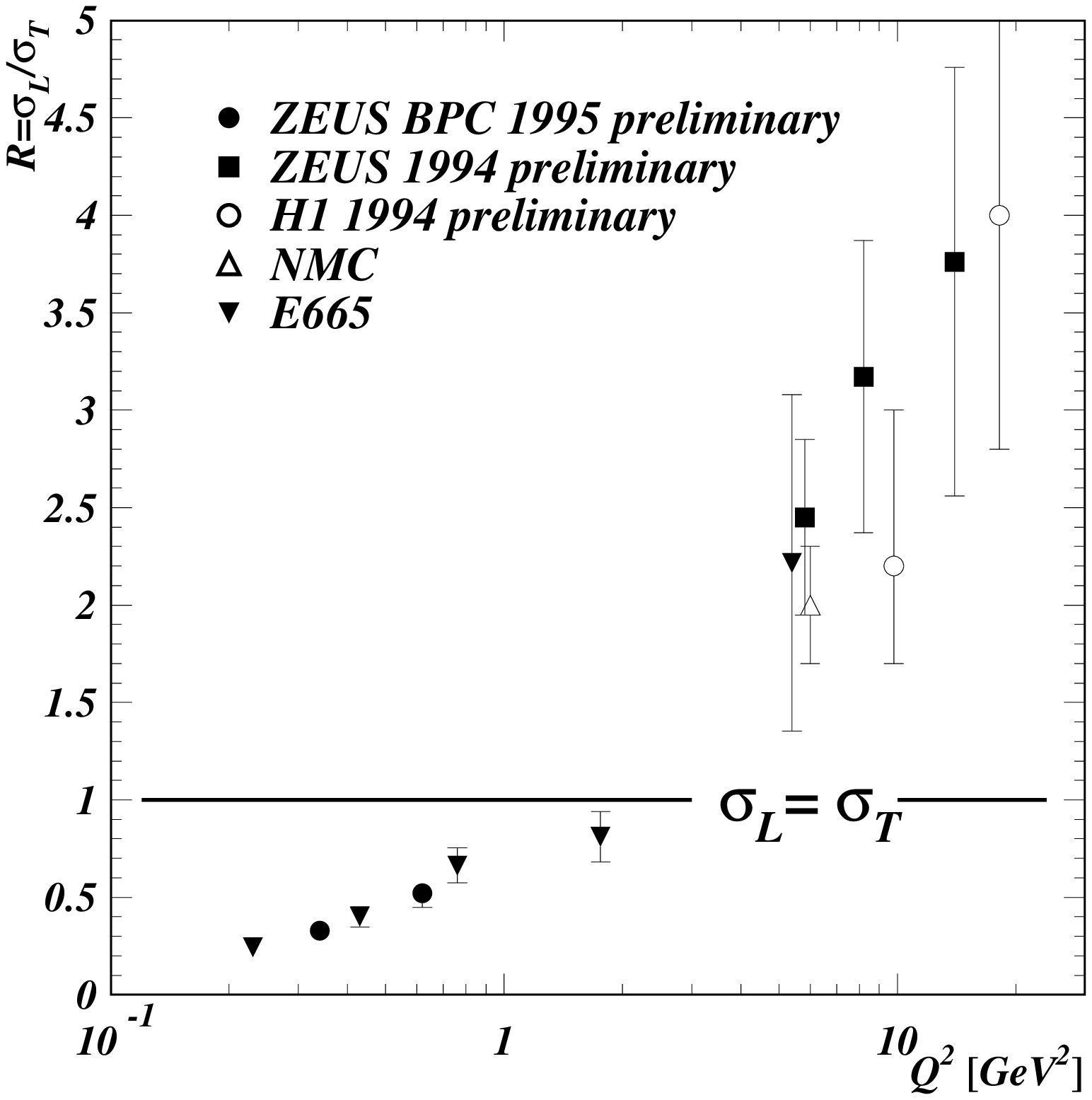}
\vspace*{-3mm}
\caption
{
\label{fig:rplot}
Values for $R$ as a function of {\qsq}
derived from the spin-density matrix element $r_{00}^{04}$
under the assumption of SCHC  \hspace*{\fill}
}
\end{center}
\end{minipage}
\vspace*{-2mm}
\end{figure}

The 
polar angle distributions measured by the ZEUS collaboration using data from their beam-pipe calorimeter and the resulting values for $r_{00}^{04}$, along with those determined from the measurements at higher {\qsq}, are shown in Fig.~\ref{fig:bpc_hel_theta}~\cite{teresa, hep97_09_031}. These results clearly show a trend toward increasing contribution from the longitudinal cross section with increasing {\qsq}. Figure~\ref{fig:rplot} 
compares these results to those from E665~\protect\cite{zfp_72_237} and H1, which support the conclusion that $R$ increases with {\qsq} as predicted in pQCD. The HERMES collaboration is also presenting measurements of $R$ in this region of {\qsq} at this workshop~\protect\cite{dueren}.

\begin{figure}[htbp]
\vspace*{-2mm}
\begin{minipage}[b]{0.56\textwidth}
Martin et al.~\cite{pr_55_4329} have calculated the {\qsq} dependence of the ratio $R$ for {\rhoz} electroproduction. Fig.~\ref{fig:r_highq2} compares their results for various assumptions of the parton distribution functions with the ZEUS and H1 measurements.
\caption
{
\label{fig:r_highq2}
Values for the ratio $R$ measured at high {\qsq} at HERA compared to
the pQCD calculations of Martin et al.~\protect\cite{pr_55_4329} using various assumptions for the parton distribution functions \hspace*{\fill}
}
\vspace*{-2mm}
\end{minipage}
\hfill
\begin{minipage}[t]{0.4\textwidth}
\raisebox{-2mm}{\includegraphics[width=\textwidth, bb= 0 25 493 491, clip=]{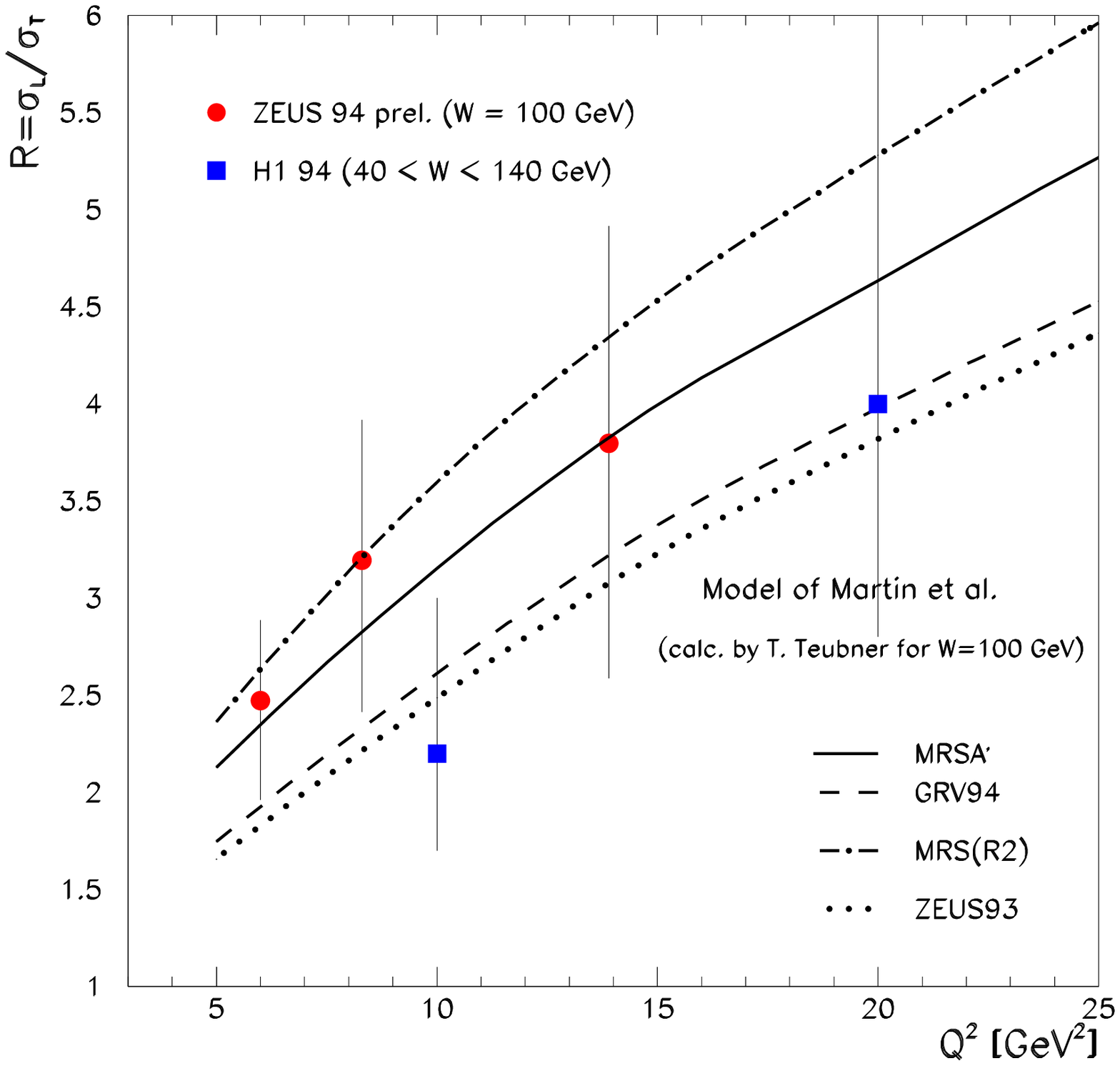}}
\end{minipage}
\end{figure}

\subsection{Spin-Density Matrix Elements at High $|t|$}
\begin{figure}[htbp]
\vspace*{-2mm}
%\begin{minipage}[t]{0.48\textwidth}
\begin{center}
\includegraphics[width=0.85\textwidth, bb= 4 0 589 314, clip=]{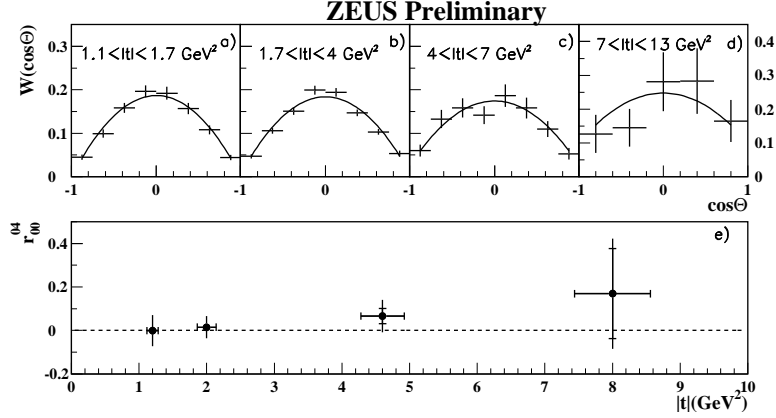}
\end{center}
%\end{minipage}
%\hfill
%\begin{minipage}[b]{0.48\textwidth}
\vskip -4mm
\caption
{
\label{fig:helhight1}
a)-d) The polar decay angle of the $\pi^+$ in {\rhoz} photoproduction in four $t$ bins measured using the ZEUS 44-m tagger, e) the value for $r_{00}^{04}$ derived from the polar angle distributions~\protect\cite{j640}.
}
%\end{minipage}
\vspace*{-2mm}
\end{figure}

The {\rhoz} photoproduction data from the ZEUS 44-m electron detector can be used to address the question of whether the ratio $R$ also rises with $|t|$. The first attempt at such an investigation is shown in Fig.~\ref{fig:helhight1}~\cite{j640}, 
where the polar angular distributions in four $t$ bins and the resulting $t$ dependence of $r_{00}^{04}$ are shown. No significant deviation from the SCHC expectation is observed, i.e. the results are consistent with no longitudinal component at high $|t|$.

\begin{figure}[htbp]
\vspace*{-2mm}
%\begin{minipage}[t]{0.48\textwidth}
\begin{center}
\includegraphics[width=0.85\textwidth, bb= 4 0 589 314, clip=]{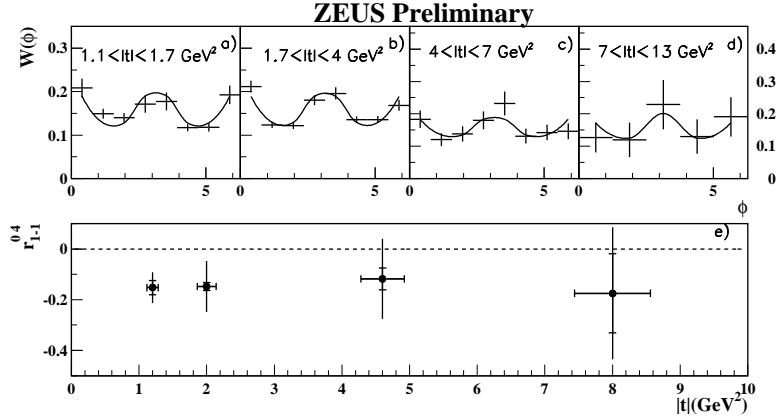}
\end{center}
%\end{minipage}
%\hfill
%\begin{minipage}[b]{0.48\textwidth}
\vskip -4mm
\caption
{
\label{fig:helhight2}
a)-d) The azimuthal decay angle of the $\pi^+$ in {\rhoz} photoproduction in four $t$ bins measured using the ZEUS 44-m tagger, and e) the value for $r_{1-1}^{04}$ derived from the azimuthal angle distributions~\protect\cite{j640}.
}
%\end{minipage}
\vspace*{-2mm}
\end{figure}

The analysis of the azimuthal angular dependence in this data set yielded a surprising result: the dependence exhibited in Fig.~\ref{fig:helhight2} 
shows a deviation from $r^{04}_{1-1}=0$, indicating a nonzero contribution from the interference of helicity-violating amplitudes. However, the interpretation of these results is complicated by the dependence of the spin-density matrix elements on the invariant mass of the pion pair in the {\rhoz} mass region. No such violation is observed for $\phi$ production in the 44-m sample; further details are available from the talk by J.~Whitmore~\cite{whitmore}.

%\clearpage
\section{Conclusions and Outlook}
\label{sec:conclusions}
Investigations into high-energy diffractive vector-meson production at HERA are yielding intriguing indications that perturbative quantum chromodynamics may provide an accurate description of these processes in specific kinematic domains. The energy dependence of the elastic {\jpsi} photoproduction cross section is inconsistent with the Pomeron exchange which describes total hadronic cross sections, but consistent with perturbative QCD calculations assuming the exchange of a colorless gluon pair. A cross section for the photoproduction of $\psi$(2S) mesons has been reported by the H1 collaboration and ZEUS has reported the observation of a signal for $\Upsilon$ photoproduction, promising interesting comparisons with pQCD calculations which invoke the vector-meson mass as the hard scale.
The pQCD calculations for {\rhoz} photoproduction predict  an energy dependence steepening with {\qsq} for {\rhoz} electroproduction, but the statistical power of the data available at present do not  allow a conclusion concerning this prediction. 

The slopes of the exponential dependence on $t$, which yield information on the spatial extent of the interaction, show a decreasing tendency both as a function of vector-meson mass and as a function of photon virtuality in the case of {\rhoz} electroproduction. The slope for {\jpsi} photoproduction is consistent with the size of the proton alone, indicating that the {\jpsi} is significantly smaller than the proton.  Several attempts to quantify an energy dependence in these slopes (shrinkage effects) have proven how difficult the measurement of such a logarithmic dependence can be. However, a recent analysis by A.~Levy excludes for {\jpsi} photoproduction the magnitude of the shrinkage effect observed in the total hadronic cross sections, encouraging a perturbative approach in describing this process.

High-statistics measurements of {\rhoz} photoproduction by the ZEUS collaboration, and isolation of a high-$|t|$ sample have allowed detailed study of the contribution of a continuum $\pi^+\pi^-$ background relative to that for resonant {\rhoz} production. The relative contribution is observed to decrease sharply with increasing $|t|$, reaching a level of less than 10\% for $|t|>1\;\gevsq$. The rates of $\phi$ and {\jpsi} photoproduction increase with increasing $|t|$ relative to that for the {\rhoz} meson, approaching values expected in the case of flavor-independent point-like photon-quark couplings. Measurements by the H1 collaboration of {\jpsi} photoproduction at high $|t|$ support the claim that the dominant interaction mechanism for $|t|>2\;\gevsq$ can be described perturbatively.

The high-energy HERA analyses of vector meson decay angle distributions yield results consistent with the $s$-channel helicity conservation (SCHC) observed at low energies. Deriving values for the ratio of longitudinal to transverse contributions to the elastic {\rhoz} cross section as a function of {\qsq} under the assumption of SCHC, the ZEUS collaboration confirmed E665 measurements identifying the range $0.2<\qsq<2\;\gevsq$ as a transition region to dominance of the longitudinal cross section, as predicted by QCD. The measurements from H1 and ZEUS for $\qsq>5\;\gevsq$ show the ratio $R$ continuing to increase, consistent with detailed pQCD calculations of the {\qsq} dependence. The ZEUS collaboration has used their high-$|t|$ sample to address the question of whether the longitudinal contribution also increases with $|t|$ in photoproduction, but found no conclusive evidence for longitudinal {\rhoz} polarization.

The future of these investigations at HERA is very promising. The multivarious tests of pQCD already available have shown that the HERA data is sensitive to the transition region in {\em each} of the variables M$_V$, {\qsq}, and $t$. With the exception of the ZEUS high-$|t|$ and $\Upsilon$ analyses, the studies described in this report are based on data recorded prior to 1996.
\begin{figure}[htbp]
\vspace*{-2mm}
\begin{minipage}[b]{0.44\textwidth}
Figure~\protect\ref{fig:lumi} shows the yearly integrated luminosities at HERA for the years 1992-97. One can conclude that an increase in statistics of a factor of five will be obtained when the presently available data have been analysed. The planned HERA luminosity upgrade program will provide a factor of about five in instantaneous luminosity, with a goal of $1\;{\rm fb}^{-1}$ by the year 2005, more than an order of magnitude greater than that recorded to date. 
\vskip 3mm
\caption
{
\label{fig:lumi}
The yearly integrated luminosity delivered by HERA during the years 1992-97.
}
\end{minipage}
\hfill
\begin{minipage}[b]{0.52\textwidth}
\vskip -4mm
\begin{center}
\includegraphics[width=\textwidth, bb= 0 26 414 485, clip=]{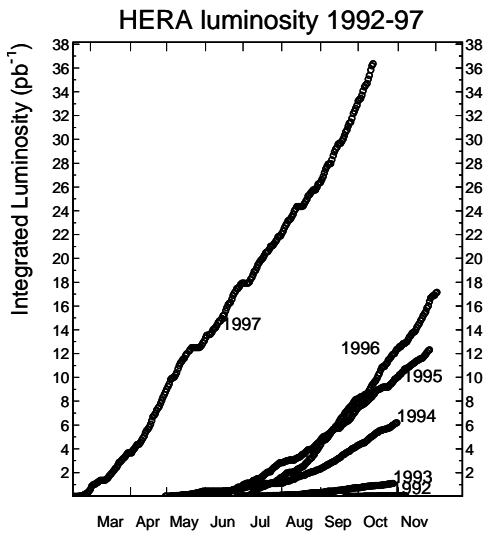}
\end{center}
\end{minipage}
\vspace*{-2mm}
\end{figure}

The statistics-limited studies such as those at high photon virtuality and those of heavy-quarkonia production can therefore be expected to improve significantly both on short and intermediate time scales. Given in addition the electron polarization program for the H1 and ZEUS experiments, which will extend the helicity analyses, we can expect many fruitful investigations of diffractive vector-meson production at high energy in the years to come.

\section*{Acknowledgements}
\addcontentsline{toc}{section}{\numberline {\mbox{ }}Acknowledgements}
\label{sec:thanks}
I would like to acknowledge useful discussions with E.~Gotsman, B.~Kopeliovich, T.~Monteiro, M.~Strikman, L.~West, and J.~Whitmore. I thank the conference organizers for their support. This work is also supported by the Bundesministerium f\"ur Bildung und Forschung.
\vspace*{5mm}
%\newpage
%\thispagestyle{empty}
%\cleardoublepage
\addcontentsline{toc}{section}{\numberline {\mbox{ }}References}
\bibliographystyle{/home/critten/tex/biblio/zeusstylem}
\bibliography{procrefs,zeuspubs,h1pubs,schriftrefs}

\end{document}